\documentclass[12pt]{article}
\usepackage{epsfig,cite}
\usepackage{amsmath}
\usepackage{amssymb}
\usepackage{hyperref}
\newcount\timecount
\newcount\hours \newcount\minutes  \newcount\temp \newcount\pmhours
\hours = \time
\divide\hours by 60
\temp = \hours
\multiply\temp by 60
\minutes = \time
\advance\minutes by -\temp
\def\hour{\the\hours}
\def\minute{\ifnum\minutes<10 0\the\minutes
            \else\the\minutes\fi}
\def\clock{
\ifnum\hours=0 12:\minute\ AM
\else\ifnum\hours<12 \hour:\minute\ AM
      \else\ifnum\hours=12 12:\minute\ PM
            \else\ifnum\hours>12
                 \pmhours=\hours
                 \advance\pmhours by -12
                 \the\pmhours:\minute\ PM
                 \fi
            \fi
      \fi
\fi
}

\def\monthname{\relax\ifcase\month 0/\or January\or February\or
   March\or April\or May\or June\or July\or August\or September\or
   October\or November\or December\else\number\month/\fi}

\def\bold#1{\setbox0=\hbox{$#1$}%
     \kern-.025em\copy0\kern-\wd0
     \kern.05em\copy0\kern-\wd0
     \kern-.025em\raise.0433em\box0 }

\def\beq{\begin{equation}}
\def\eeq{\end{equation}}

\def\gev{{\rm \, Ge\kern-0.125em V}}
\def\tev{{\rm \, Te\kern-0.125em V}}
\def\gyr{{\rm \, G\kern-0.125em yr}}
\def\ohsq{\Omega_{\chi} h^2}

\def\nl{\hfill\nonumber\\&&}
\def\nnl{\hfill\nonumber\\}

\def\gappeq{\mathrel{\rlap {\raise.5ex\hbox{$>$}}
{\lower.5ex\hbox{$\sim$}}}}
\def\lappeq{\mathrel{\rlap{\raise.5ex\hbox{$<$}}
{\lower.5ex\hbox{$\sim$}}}}
\def\Toprel#1\over#2{\mathrel{\mathop{#2}\limits^{#1}}}

\def\schi{\widetilde \chi}        

\def\sel{{\widetilde e}}

\def\stau{\widetilde \tau}
\def\stop{\widetilde t}

\def\m12{m_{1\!/2}}

\def\mstau{m_{\tilde{\ell}_1}}

\def\PL{{Phys.~Lett.} }
\def\PR{{Phys.~Rev.} }

\def\stau{\tilde{\tau}}
\def\mstau{m_{\tilde{\tau}}}

\def\mpl{M_{P}}

\def\bea{\begin{eqnarray}}
\def\eea{\end{eqnarray}}

\def\mplr{\overline{M_{P}}}
\def\mgut{M_{GUT}}
\def\calh{\mathcal{H}}
\def\hfiv{h_\mathbf{\overline{5}}}
\def\hten{h_\mathbf{10}}
\def\mfiv{m_\mathbf{\overline{5}}}
\def\mfivl{m_{\mathbf{\overline{5}},1}}
\def\mten{m_\mathbf{10}}
\def\mtenl{m_{\mathbf{10},1}}
\def\afiv{A_\mathbf{\overline{5}}}
\def\aten{A_\mathbf{10}}

\mathchardef\mhyphen="2D

\DeclareMathOperator{\Tr}{Tr}

\begin{document}
\begin{titlepage}
\pagestyle{empty}
\baselineskip=21pt
\rightline{CERN-PH-TH/2010-065}
\rightline{UMN--TH--2840/10}
\rightline{FTPI--MINN--10/09}
\vskip 0.2in
\begin{center}
{\large{\bf What if Supersymmetry Breaking Unifies beyond the GUT Scale?}}
\end{center}
\begin{center}
\vskip 0.2in
{\bf John~Ellis}$^1$, {\bf Azar Mustafayev}$^{2}$ and {\bf Keith~A.~Olive}$^{2}$

\vskip 0.1in

{\it
$^1${TH Division, PH Department, CERN, CH-1211 Geneva 23, Switzerland}\\
$^2${William I.~Fine Theoretical Physics Institute, \\
University of Minnesota, Minneapolis, MN 55455, USA}\\
}

\vskip 0.2in
{\bf Abstract}
\end{center}
\baselineskip=18pt \noindent

We study models in which soft supersymmetry-breaking parameters of the MSSM become universal
at some unification scale, $M_{in}$, above the GUT scale, $\mgut$.  We assume
that the scalar masses and gaugino masses have common values, $m_0$ and $m_{1/2}$
respectively,  at
$M_{in}$. We use the renormalization-group equations of the minimal supersymmetric SU(5) GUT
to evaluate their evolutions down to $\mgut$, studying their dependences on the unknown
parameters of the SU(5) superpotential. After displaying some generic examples of the
evolutions of the soft supersymmetry-breaking parameters, we discuss the effects on physical
sparticle masses in some specific examples. We note, for example, that near-degeneracy
between the lightest neutralino and the lighter stau is progressively disfavoured as $M_{in}$
increases. This has the consequence, as we show in $(m_{1/2}, m_0)$ planes for several different values of $\tan \beta$, 
that the stau coannihilation region shrinks as $M_{in}$ increases, and
we delineate the regions of the $(M_{in}, \tan \beta)$ plane where it is absent altogether.
Moreover, as $M_{in}$ increases, the focus-point region recedes to larger values of $m_0$
for any fixed $\tan \beta$ and $m_{1/2}$. We conclude that the regions of the $(m_{1/2}, m_0)$
plane that are commonly favoured in phenomenological analyses tend to disappear at large $M_{in}$.


\vfill
\leftline{CERN-PH-TH/2010-065}
\leftline{March 2010}
\end{titlepage}

\section{Introduction}

The minimal supersymmetric extension of the Standard Model (MSSM) has over
100 free parameters, most of them associated with the breaking of 
supersymmetry~\cite{mssm,Baer:book,Drees:book}.
Three classes of soft supersymmetry-breaking parameters are generally considered,
the scalar masses $m_0$, the gaugino masses $m_{1/2}$ and the trilinear scalar
couplings $A_0$, which are often assumed to be universal at some high input scale.
Universality before renormalization of the $m_0$ parameters for different sfermions 
with the same electroweak quantum numbers is motivated by the upper limits on
flavour-changing neutral interactions, and specific Grand Unified Theories (GUTs)
suggest universality relations between squarks and sleptons~\cite{EN}. Simple GUTs also
favour universality before renormalization for the gaugino masses $m_{1/2}$,
and universality is also a property of minimal supergravity (mSUGRA) models,
which, however, also predict additional relations that we do not discuss here~\cite{BIM,bfs,vcmssm}.
We refer to the scenario with universal $m_0$, $m_{1/2}$ and $A_0$ as the
constrained MSSM (CMSSM), and its parameter space has been explored 
extensively~\cite{funnel,cmssm,efgosi,cmssmnew,cmssmmap,like1,like2}.

There is, however, one important question: at what renormalization scale $M_{in}$ are $m_0$
and $m_{1/2}$ actually universal? The obvious possibility, and the one has been studied most
frequently, is that universality applies at the same GUT scale, $\mgut$, as coupling constant
universality. In this case, the
density of cold dark matter (assumed here to be composed mainly of the lightest neutralino, 
$\schi_1$, hereafter called $\chi$)~\cite{EHNOS} 
is larger than the range favoured by WMAP~\cite{WMAP} and other experiments in generic regions 
of the $(m_{1/2}, m_0)$ plane, and is compatible with
WMAP only in narrow strips that are either close to the boundary where $\chi$
ceases to be the lightest sparticle -- the stau~\cite{stauco} or stop~\cite{dm:stop} coannihilation strips 
-- or close to the LEP2 chargino bound~\cite{LEPsusy} -- the bulk region~\cite{EHNOS,cmssm} 
-- or where there is no electroweak symmetry breaking -- the focus-point region~\cite{focus} 
-- or in a rapid-annihilation funnel (or A-funnel)~\cite{funnel}.

However, it is not necessarily the case that $M_{in} = \mgut$, 
since supersymmetry breaking might arise at some scale either
below or even above $\mgut$, and both possibilities have been studied in the literature.
For example, as $M_{in}$ is decreased below $\mgut$, the differences between the
renormalized sparticle masses diminish and the regions of the $(m_{1/2}, m_0)$
planes that yield the appropriate density of cold dark matter move away from the boundaries~\cite{EOS06}.
Eventually, for small $M_{in}$, the coannihilation and focus-point regions of the conventional
GUT-scale CMSSM merge. Finally, for very small $M_{in}$ they disappear 
entirely, and the relic $\chi$
density falls before the WMAP range everywhere in the $(m_{1/2}, m_0)$ plane, except for
very large values of $m_{1/2}$ and $m_0$.

What happens to the supersymmetric parameter space and sparticle phenomenology
when $M_{in} > \mgut  \sim 2 \times 10^{16}$~GeV? 
Here, we consider values of $M_{in}$ ranging up to the
reduced Planck mass $\mplr \equiv \mpl /\sqrt{2\pi} \sim 2.4 \times 10^{18}$~GeV.
Generically, increasing
$M_{in}$ increases the renormalization of the sparticle masses which tends in turn to increase
the splittings between the physical sparticle masses~\cite{pp}. As we discuss in more detail below,
this in turn has the effect of increasing the relic density in much of the $(m_{1/2}, m_0)$
plane. As a consequence, the coannihilation strip is squeezed to lower values of $m_{1/2}$~\footnote{For 
a previous example of this phenomenon, see Fig. 5 of~\cite{Calibbi} or Fig.~3 of~\cite{CEGLR}.},
particularly for $\tan \beta \sim 10$, and
even disappears as $M_{in}$ increases. At the same time, the focus-point strip often moves out
to ever larger values of $m_0$. There are also changes in the impacts of important constraints
such as $g_\mu - 2, b \to s \gamma$ and $m_h$, which we also discuss below. The general
conclusion is that the supersymmetric landscape would look rather different for 
$M_{in} > 10^{17}$~GeV from the CMSSM in which the universality scale $M_{in} = \mgut$.
The allowed region of parameter space that survives longest is the rapid-annihilation
funnel at large $m_{1/2}$ and $\tan \beta$, which is compatible with the $m_h$ and
$b \to s \gamma$ constraints. In the CMSSM, the funnel region also requires large
$m_0$ and would make a contribution to $g_\mu - 2$ that is
too small to explain the experimental discrepancy with Standard Model calculations
based on low-energy $e^+ e^-$ data. However, as we shall show, for large $M_{in}$,
the funnel region extends to low $m_0$ (including $m_0 = 0$) and in some cases
will be compatible with the $g_\mu - 2$ measurements.

We underline that there are some potential ambiguities in these conclusions.
We use for our analysis above $\mgut$  the particle content and the 
renormalization-group equations (RGEs)
of the minimal SU(5) GUT~\cite{pp,others}, primarily for simplicity and so as to minimize the number of
additional parameters to be explored: for a recent review of this sample model and its
compatibility with experiment, see~\cite{Senjanovic:2009kr}.
Even in this simplest GUT, there are two couplings in the
SU(5) superpotential that make potentially significant contributions to the RGE running
but are poorly constrained. We explore their impacts on our results, and find that 
one of the couplings could have noticeable effects
in the focus-point region, if it is large. Secondly, we are aware that the minimal SU(5) model is
surely inadequate; for example, it does not include neutrino masses. The effects of a
minimal seesaw sector on the GUT RGEs~\cite{casa} and on the relic density~\cite{Calibbi,CEGLR,kov} has been considered elsewhere: they also
may be small if the neutrino Dirac Yukawa couplings are not large. However, the
minimal SU(5) GUT also has issues with proton decay via dimension-five operators, which 
may be alleviated if the GUT triplet Higgs particles are relatively heavy, as would
happen if the associated SU(5) superpotential coupling were large~\cite{protdec}. 
We therefore
consider this option as the default in our analysis. One could in principle consider
non-minimal GUT models in which dimension-five proton decay is suppressed by some
other mechanism, but the exploration of such models would take us too far from our
objective here.

The layout of this paper is as follows. In Section~\ref{sec:su5}, we recall the superpotential of the
minimal SU(5) GUT and the corresponding RGEs for the soft supersymmetry-breaking
parameters. Here, we give some examples of the RGE running, assuming universality 
at some high scale $M_{in} > \mgut$. We also give examples of the dependences of
physical sparticle masses on $M_{in}$, illustrating features that are important for
understanding qualitatively the dependences of features in the $(m_{1/2}, m_0)$ plane.
Several of these are displayed in Section~\ref{sec:plane}, for representative values of $M_{in}$ and
default values of the unknown SU(5) superpotential parameters. As already mentioned,
two of the striking features in these planes are the disappearance of the
coannihilation strip and the movement of the focus-point strip to larger $m_0$ as
$M_{in}$ increases. We discuss in Section~\ref{sec:tanb} the sensitivities of these features to the choices
of the SU(5) superpotential parameters, showing that the disappearance of the
coannihilation strip is relatively model-independent, whereas the movement of the
focus-point strip is more model-dependent. Finally, in Section~\ref{sec:concl} we summarize our results
and draw some conclusions for the generalization of the standard CMSSM with
$M_{in} = \mgut$ to different values of $M_{in}$.

\section{The Minimal SU(5) GUT Superpotential and RGEs}
\label{sec:su5}

In the SU(5) GUT, the $\hat{D}^c_i$ and $\hat{L}_i$ superfields of the MSSM reside in the $\bf{\overline{5}}$
representation, $\hat{\phi}_i$, while the $\hat{Q}_i,\ \hat{U}^c_i$ and $\hat{E}^c_i$ superfields are in the $\bf{10}$ representation,
$\hat{\psi}_i$. 
In the minimal scenario, one introduces a single SU(5) adjoint Higgs
multiplet $\hat{\Sigma}(\bf{24})$, and the two Higgs doublets of the MSSM, $\hat{H}_d$ and $\hat{H}_u$
are extended to five-dimensional SU(5) representations $\hat{\calh}_1(\bf{\overline{5}})$ and 
$\hat{\calh}_2(\bf{5})$ respectively.
The minimal renormalizable superpotential for this model is
\bea
W_5 &=& \mu_\Sigma \Tr\hat{\Sigma}^2 + \frac{1}{6}\lambda'\Tr\hat{\Sigma}^3
 + \mu_H \hat{\calh}_{1\alpha} \hat{\calh}_2^{\alpha} 
 + \lambda \hat{\calh}_{1\alpha}\hat{\Sigma}^{\alpha}_{\beta} \hat{\calh}_2^{\beta} \nl
 +({\bf h_{10}})_{ij} \epsilon_{\alpha\beta\gamma\delta\zeta}
   \hat{\psi}^{\alpha\beta}_i \hat{\psi}^{\gamma\delta}_j \hat{\calh}_2^{\zeta} 
 +({\bf h_{\overline{5}}})_{ij} \hat{\psi}^{\alpha\beta}_i \hat{\phi}_{j\alpha} \hat{\calh}_{1\beta}
\label{W5}
\eea
where Greek letters denote SU(5) indices, $i,j=1..3$ are generation indices and $\epsilon$ is
the totally antisymmetric tensor with $\epsilon_{12345}=1$.
This simple model predicts (approximately) correctly the observed ratio of the $\tau$ and
$b$ quark masses, but the corresponding predictions for the lighter charged-lepton and
charge -1/3 quark masses are at best qualitatively successful. It is possible to add to
(\ref{W5}) terms that are non-renormalizable quartic and of higher order in the Higgs fields
that could rectify these less successful predictions: such terms would not contribute to
the RGEs and low-energy observables that we study. 
In this paper, we will work in the third-generation-dominance scheme 
where Yukawas of first two generations are neglected,
{\it i.e.}, we assume ${\bf h_{{\overline{5}},10}} \sim \left({\bf h_{{\overline{5}},10}}\right)_{33} \equiv h_{\mathbf{\overline{5}},\mathbf{10}}$.

We work in a vacuum that breaks SU(5) $\to$ SU(3) $\times$ SU(2) $\times$ U(1), in which
$\langle \hat{\Sigma} \rangle = v_{24} {\rm Diag}(2, 2, 2, -3, -3)$ and the GUT gauge bosons
acquire masses $M_{X, Y} = 5 g_{GUT} v_{24}$. The fine-tuning condition 
$\mu_\Sigma - 3 \lambda v_{24} = {\cal O}(M_Z)$ must be imposed in order to obtain the gauge
hierarchy, in which case the triplet Higgs states have masses $M_{H_3} = \lambda v_{24}/g_{GUT}$.
The amplitude for proton decay via a dimension-five operator $\propto 1/M_{H_3}$, and so
is relatively suppressed for large $\lambda$. However, the amplitude also depends on
other model parameters, so it is difficult to quantify this argument, which would in any case
be avoided in suitable non-minimal SU(5) models. In this paper we compare results
for the values $\lambda = 1$ and 0.1, treating the former as our default value.

The RGEs for the Yukawa couplings in the superpotential (\ref{W5}) that are applicable
between $M_{in}$ and $\mgut$ are:
\bea
\frac{d \hfiv}{dt} & = & 
  \frac{\hfiv}{16 \pi^2} \left[ 5\hfiv^2 + 48\hten^2 + \frac{24}{5} \lambda^2 
  - \frac{84}{5} g_5^2 \right], \\
\frac{d \hten}{dt} & = & 
  \frac{\hten}{16 \pi^2} \left[ 144\hten^2 + 2\hfiv^2 + \frac{24}{5} \lambda^2 
  - \frac{96}{5} g_5^2 \right], \\
\frac{d \lambda}{dt} & = & 
  \frac{\lambda}{16 \pi^2} \left[ 48\hten^2 + 2\hfiv^2 + \frac{53}{5} \lambda^2 
  + \frac{21}{20} \lambda'^2 - \frac{98}{5} g_5^2 \right], \\
\frac{d \lambda'}{dt} & = & 
  \frac{\lambda'}{16 \pi^2} \left[ 3 \lambda^2 + \frac{63}{20} \lambda'^2
  - 30 g_5^2 \right], 
\label{RGEYukawa}
\eea
where $g_5$ is the SU(5) gauge coupling above the GUT scale.
We note that the Yukawa coupling $\lambda$ contributes directly to the
RGEs for $\hfiv$ and $\hten$ while $\lambda'$ contributes indirectly through its effect on the running of $\lambda$. In most analyses of the CMSSM, 
$\hfiv$ and $\hten$ are chosen at the GUT scale so
as to reproduce the measured $t$ and $b$ masses. Equations (4) and (5) tell us that the input
values required at $M_{in}$ depend on $\lambda$. Other quantities renormalized
by $\hfiv$ and $\hten$, 
such as the third-generation tenplet mass $\mten$ and fiveplet mass $\mfiv$,
will thereby acquire some indirect dependence on $\lambda$. 

The RGEs for the most relevant soft supersymmetry-breaking squared scalar masses between $M_{in}$ and $\mgut$ are:
\bea
\frac{d \mfiv^2}{dt} & = & 
  \frac{1}{8 \pi^2} \left[ 2\hfiv^2 \{ m^2_{\calh_1} + \mten^2 + \mfiv^2 + \afiv^2 \} 
   - \frac{48}{5} g_5^2 M_5^2 \right], \label{RGEmfiv}\\
\frac{d \mten^2}{dt} & = & 
  \frac{1}{8 \pi^2} \biggl[ 48\hten^2 \{ m^2_{\calh_2} + 2\mten^2 + \aten^2 \} \biggr.\nl
   \left.\qquad +\hfiv^2 \{ m^2_{\calh_1} + \mten^2 + \mfiv^2 + \afiv^2 \} 
   - \frac{72}{5} g_5^2 M_5^2 \right], \label{RGEmten}\\
\frac{d m_{\calh_1}^2}{dt} & = & 
  \frac{1}{8 \pi^2} \biggl[ 2\hfiv^2 \{ m^2_{\calh_1} + \mten^2 + \mfiv^2 + \afiv^2 \} \biggr.\nl
   \left. \qquad +\frac{24}{5} \lambda^2 \{ m^2_{\calh_1} + m^2_{\calh_2} + m_{\Sigma}^2 + A_\lambda^2 \} 
   - \frac{48}{5} g_5^2 M_5^2 \right],\\
\frac{d m_{\calh_2}^2}{dt} & = & 
  \frac{1}{8 \pi^2} \biggl[ 48\hten^2 \{ m^2_{\calh_2} + 2\mten^2 + \aten^2 \} \biggr.\nl
   \left.\qquad +\frac{24}{5} \lambda^2 \{ m^2_{\calh_1} + m^2_{\calh_2} + m_{\Sigma}^2 + A_{\lambda}^2 \} 
   - \frac{48}{5} g_5^2 M_5^2 \right],  \label{RGEh2}\\
\frac{d m_{\Sigma}^2}{dt} & = & 
  \frac{1}{8 \pi^2} \left[ \frac{21}{20} \lambda'^2 \{ 3 m^2_{\Sigma} + A_{\lambda'}^2 \} 
   +\lambda^2 \{ m^2_{\calh_1} + m^2_{\calh_2} + m_{\Sigma}^2 + A_\lambda^2 \} 
   - 20 g_5^2 M_5^2 \right].
\eea
The RGEs for the first two generations' sfermion masses $\mtenl$ and $\mfivl$ contain only gauge-gaugino parts, which are identical
to their third-generation counterparts~(\ref{RGEmfiv}, \ref{RGEmten}). 
We note that all the scalar masses-squared are strongly renormalized by $g_5$, some of
whose implications we shall discuss later. 

We see that the extra GUT superpotential couplings $\lambda, \lambda'$ do not affect directly the
evolutions of the soft supersymmetry-breaking masses of the matter multiplets. However, the
value of $\lambda$ may have important effects on the evolutions of the 
soft supersymmetry-breaking masses of the fiveplet Higgs multiplets.
On the other hand, the value
of $\lambda'$ impacts directly only the adjoint Higgs multiplet, which then contributes in turn to the
evolutions of the masses of the fiveplet Higgs multiplets. This leads us to expect that our results
should be relatively insensitive to $\lambda'$. However, we do expect them to be sensitive to $\lambda$,
particularly as one approaches the focus-point region at large $m_0$, since it lies close to
the boundary of consistent electroweak symmetry breaking (EWSB).
We do not discuss here the RGEs for the trilinear soft supersymmetry-breaking parameters $A_i$: the
patterns of their dependences on $\lambda$ and $\lambda'$ are similar to the RGEs for the
scalar masses-squared. The complete set of RGEs for minimal SU(5) can be found in Ref.~\cite{pp,Baer:2000gf} with
appropriate changes of 
notation\footnote{Our sign convention for the $A$ terms is the same as in 
Ref.~\cite{Baer:book,Drees:book,pp}, but opposite from that in Ref.\cite{Baer:2000gf} 
and in the {\tt ISAJET} interface~\cite{isajet}.}$^,$\footnote{
Because of different assumptions, our results cannot be compared directly with those of Ref.[22], 
though there are similarities.}.

The model is specified by the following set of parameters
\beq
m_0,\ m_{1/2},\ A_0,\ M_{in},\ \lambda,\ \lambda',\ \tan \beta,\ sign(\mu)
\eeq
where the trilinear superpotential Higgs couplings, $ \lambda,\ \lambda'$, 
are specified at $Q=\mgut$. 
Assuming universality at a unification scale $M_{in}$, we impose
\bea
\mfivl=\mtenl=\mfiv=\mten=m_{\calh_1}=m_{\calh_2}=m_{\Sigma} &\equiv & m_0, \nnl
\afiv=\aten=A_{\lambda}=A_{\lambda'} &\equiv & A_0, \nnl
M_5 &\equiv & m_{1/2}.
\eea
and evolve all parameters to $\mgut$ using the SU(5) RGEs mentioned earlier. 
Clearly, the CMSSM is realized by setting $M_{in}=\mgut$, and in this case, the couplings
$\lambda$ and $\lambda'$ have no effect on the low-energy spectrum.
Bilinear superpotential parameters
$\mu_H,\ \mu_{\Sigma}$ and corresponding soft supersymmetry breaking 
terms decouple from the rest of RGEs and therefore are omitted. 
The transition to the MSSM is done at $\mgut$ via the following matching conditions:
\bea
g_1=g_2=g_3 = g_5 \, , && h_t =4\hten ,\nnl
M_1=M_2=M_3 = M_5 \, , && \nnl
m_{D_1}^2 = m_{L_1}^2 = \mfivl^2 \, , && m_{Q_1}^2=m_{U_1}^2=m_{E_1}^2 = \mtenl^2 , \nnl
m_{D_3}^2 = m_{L_3}^2 = \mfiv^2 \, , && m_{Q_3}^2=m_{U_3}^2=m_{E_3}^2 = \mten^2 , \nnl
m_{H_d}^2 = m_{\calh_1}^2 \, , && m_{H_u}^2 = m_{\calh_2}^2 .
\eea
Note that we do not impose $b-\tau$ Yukawa unification at $\mgut$; 
although exact unification is possible in the MSSM, it is not guaranteed over the entire parameter space, 
and in a GUT non-renormalizable operators
of the type needed to modify the `bad' relations for the first two generations 
and/or modify the proton decay
predictions~\cite{protdec,nro} may also modify $m_b/m_\tau$. 
In the models discussed below, the ratio of $h_b/h_\tau$ is similar to that found in
the CMSSM.  Here, we typically find that $h_b/h_\tau \simeq$ 0.65 -- 0.75 for $\tan \beta = 10$,
and somewhat lower  values $h_b/h_\tau \simeq 0.5 - 0.6$ 
for $\tan \beta = 55$. It is not possible in these models to force Yukawa coupling unification
by choosing a suitable input value of $h_b/h_\tau$~\cite{Dedes:1997md}.
Therefore, we use the following matching condition for the down-sector Yukawa couplings:
\beq
(h_b+h_\tau)/2=\hfiv/\sqrt{2} \ . 
\eeq
The matching conditions for the
$A$ terms are the same as those of the corresponding Yukawa couplings.

We use for our calculations the program {\tt SSARD}~\cite{ssard}, 
which allows the computation of sparticle spectrum on the basis of
2-loop RGE evolution for the MSSM and 1-loop evolution
for minimal SU(5). 
We define $\mgut$ as the scale where $g_1=g_2$, so 
$\mgut \simeq 1.5\times10^{16}$~GeV, but its exact value depends on the location in the parameter space. 
We also performed cross-checks 
with {\tt ISAJET~7.80}~\cite{isajet} modified to include SU(5) running above $\mgut$, 
and found results in very good agreement~\footnote{For a
comparison of the {\tt SSARD} and {\tt ISAJET} codes, see Ref.~\cite{Battaglia:2001zp}.}.

\begin{figure}[ht!]
\epsfig{file=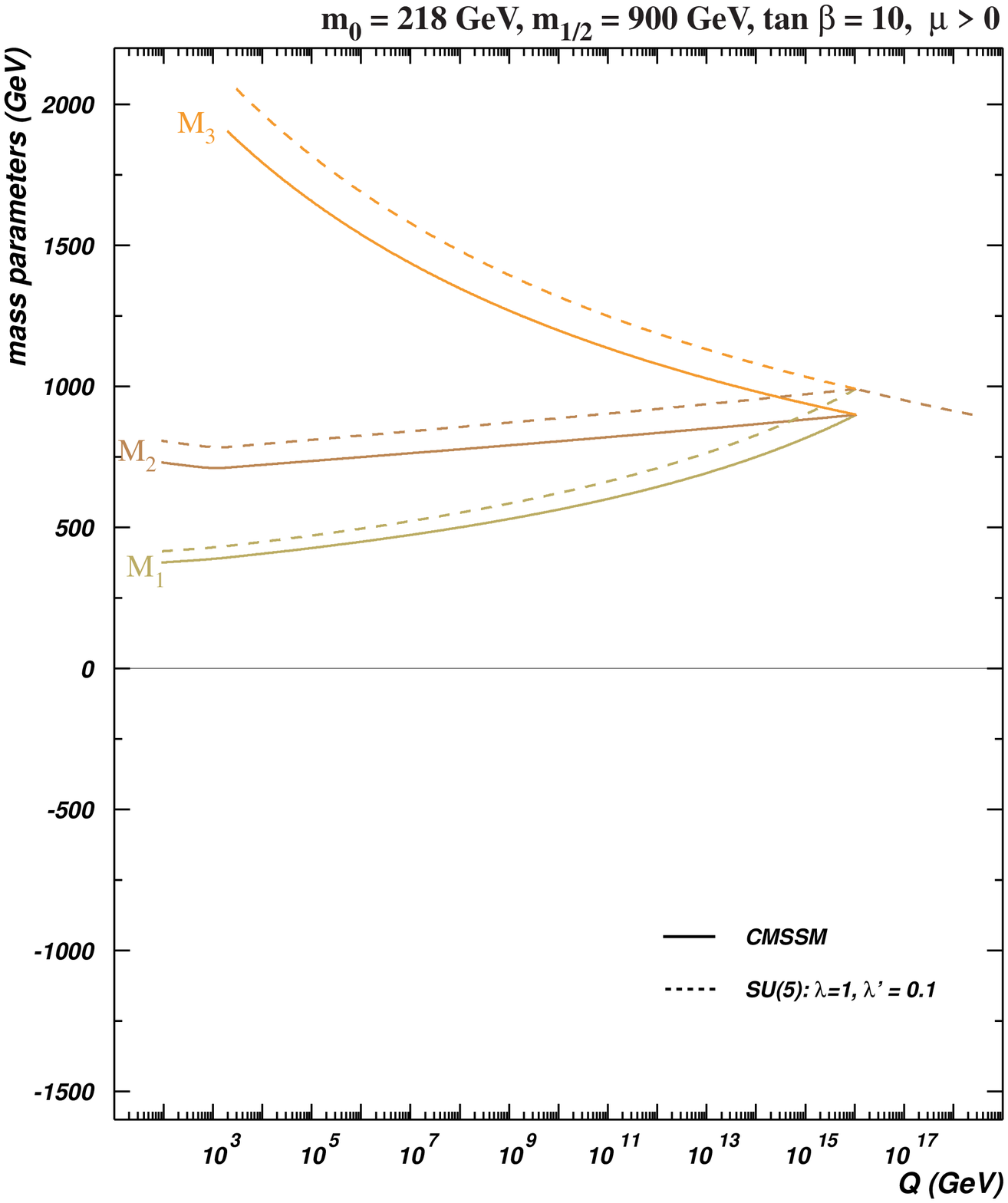,width=7cm}
\hskip .2in
\epsfig{file=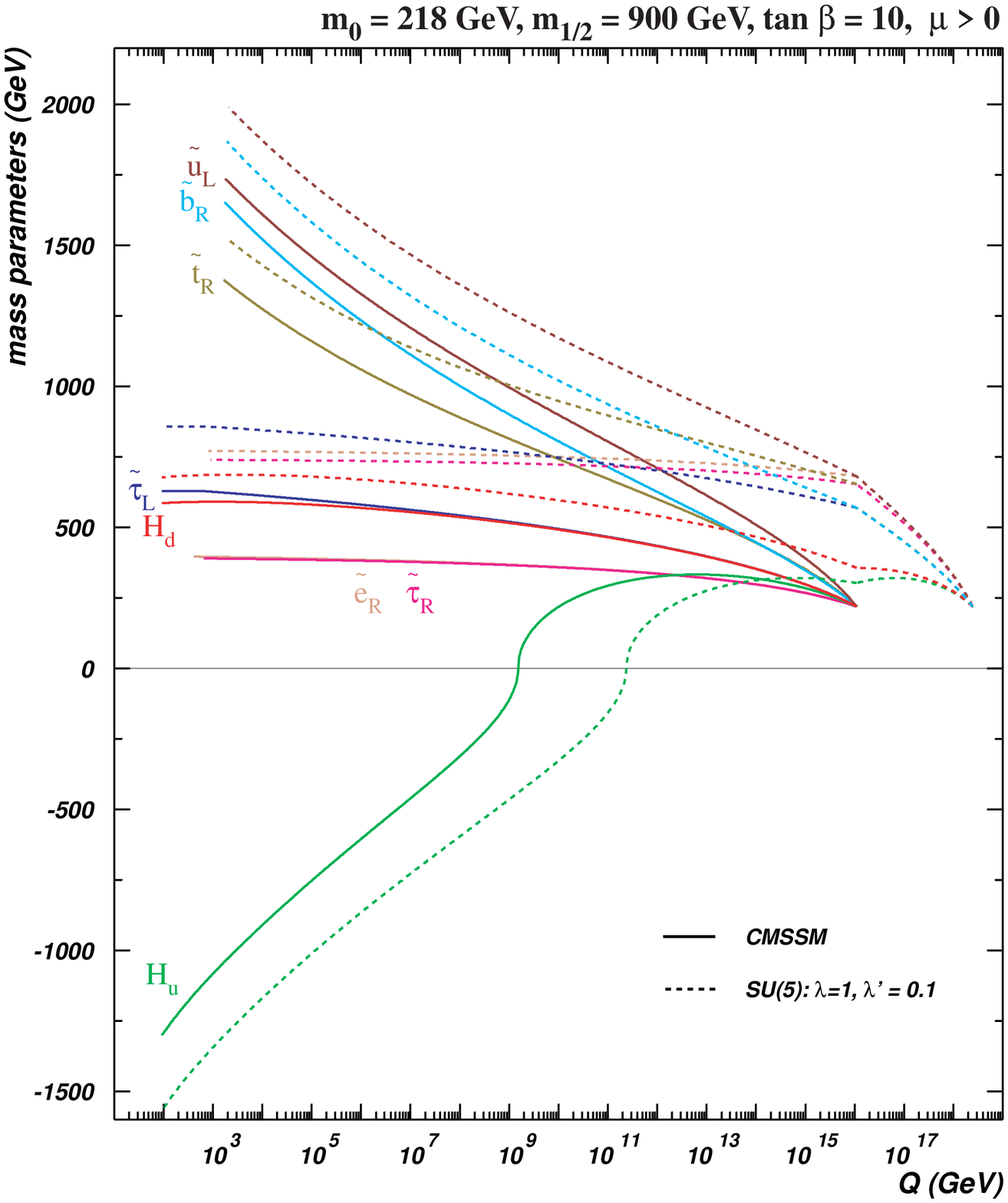,width=7cm} \\
\epsfig{file=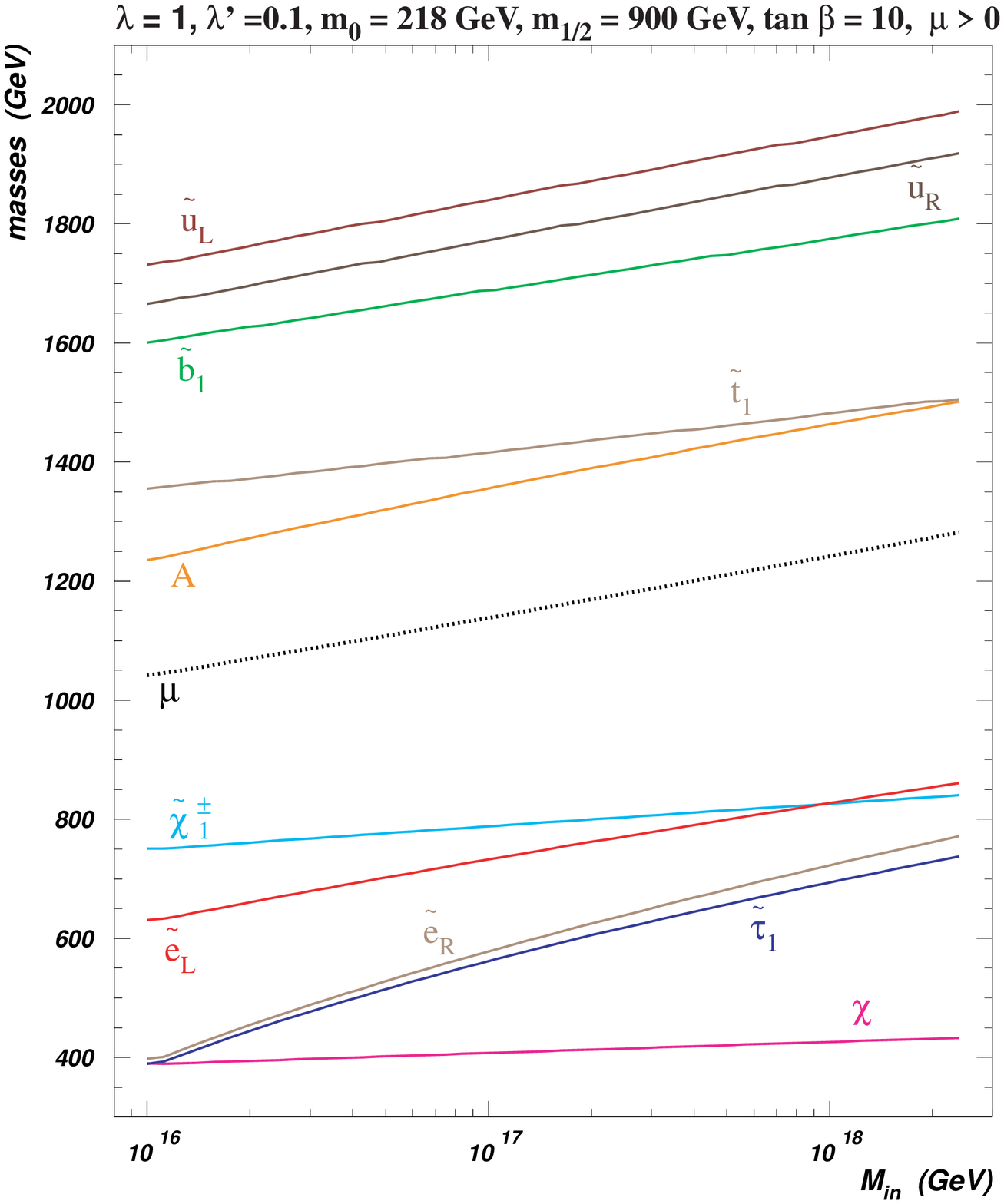,width=7cm}
\hskip .2in
\epsfig{file=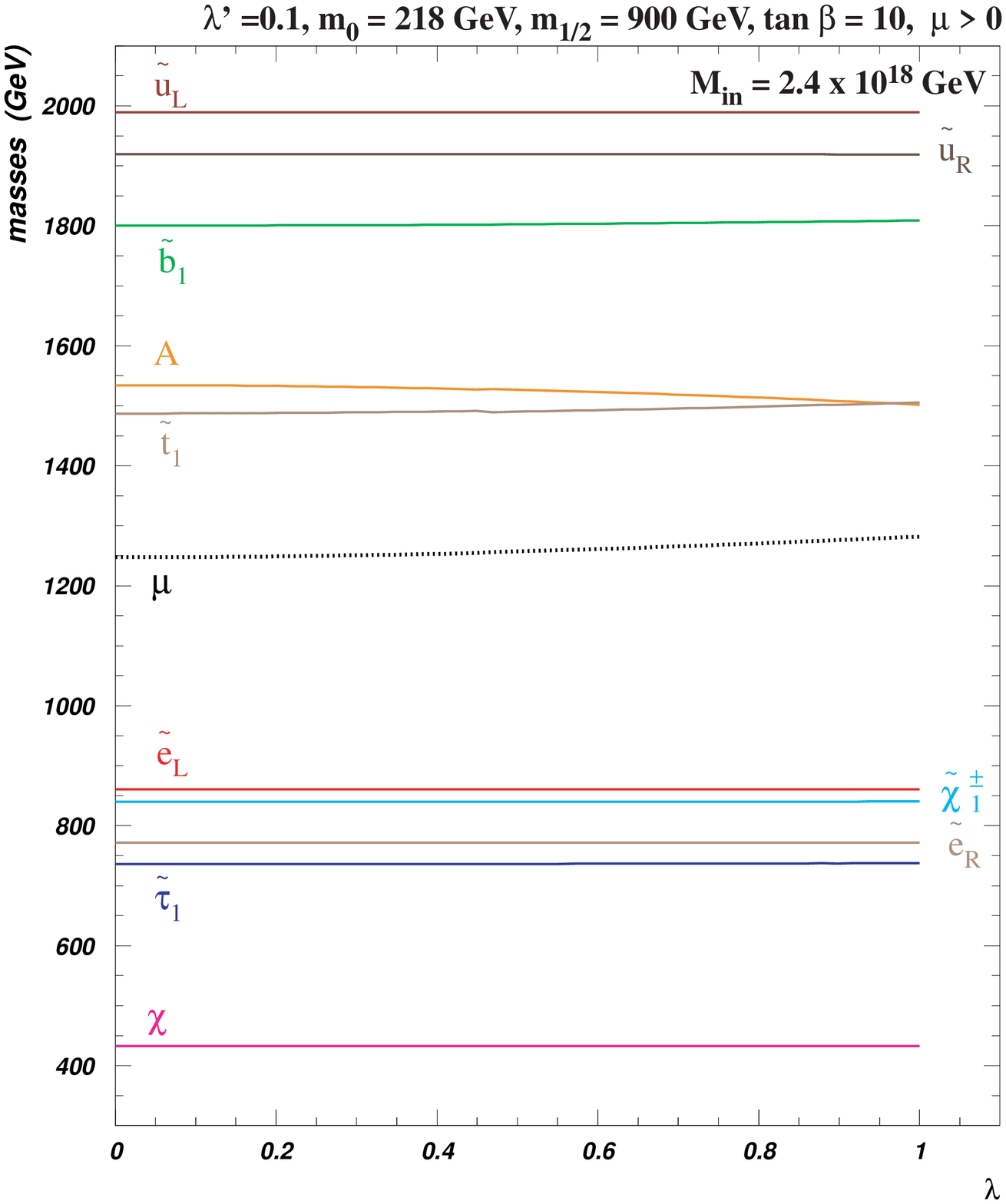,width=7.3cm}
\caption{\it
For a coannihilation point
with $m_0 = 218$~GeV, $m_{1/2} = 900$~GeV, $A_0 = 0$, $\tan \beta = 10$, and $\lambda' = 0.1$.
Top Left: the evolutions of the gaugino mass parameters 
with the choices $M_{in} = \mgut$ (solid lines) and $\mplr = 2.4 \times 10^{18}$~GeV
(dashed lines), assuming $\lambda = 1$;
Top Right: the evolutions of the soft supersymmetry-breaking scalar mass parameters
with $M_{in} = \mgut$ (solid lines) and $\mplr$ (dashed lines), assuming $\lambda = 1$;  
Bottom Left: the dependences of the physical sparticle/Higgs masses on $M_{in}$ 
 assuming $\lambda = 1$; 
Bottom Right: the dependences of the sparticle/Higgs masses on $\lambda$
 assuming $M_{in} = \mplr$. }
\label{fig:RGEsCA10}
\end{figure}

As a first illustration of the effects of the RGEs between $M_{in}$ and $\mgut$,
we show in the top panels of Fig.~\ref{fig:RGEsCA10} a comparison between the
running of the soft masses: we plot $sign(m^2_i)\sqrt{|m^2_i|}$ for scalars and $M_j$ for gauginos. 
Solid lines represent the CMSSM and dashed lines the SU(5) model 
specified at $M_{in}= 2.4 \times 10^{18}$~GeV with the same
$\tan \beta = 10$ and $m_{1/2} = 900$~GeV, $m_0 = 218$~GeV and $A_0 = 0$. 
This point is chosen because it lies
near the tip of the WMAP coannihilation strip for $\tan \beta = 10$ in the CMSSM and is similar to
benchmark point H of~\cite{Battaglia:2001zp}.
There is some quantitative change in the
evolution of gaugino masses $M_i$, but their qualitative behaviour is similar in the two models.
On the other hand, the Higgs mass-squared parameters are significantly renormalized by 
$g_5$, since the $m_{1/2}$ gauge-gaugino term dominates in the SU(5) RGEs.
In the case of SU(5), the large value of $M_5$ initially causes $\calh_1$ and $\calh_2$ masses 
to evolve upward, but ultimately
the $\lambda^2$ term takes over and reverses the direction of evolution. 
As a result, the $\calh_{1,2}$ masses at $\mgut$ are not too large compared to $m_0^2$, 
and $m_{\calh_1}^2 > m_{\calh_2}^2$ due to the relative strength of the $\hfiv^2$ and $\hten^2$ terms.
The running of $m_{\calh_1}^2$ and $m_{\calh_2}^2$ between $M_{in}$ and $\mgut$
evidently yields $m_{H_u}^2 \ne m_{H_d}^2 \ne m_0^2$ at the GUT scale. This is equivalent 
to the structure found in non-universal Higgs mass models (NUHM)~\cite{nonu,nuhm}.
As a result, the following combination of soft 
supersymmetry-breaking parameters~\cite{Martin:1993zk}: 
\begin{eqnarray}
S &\equiv&  m_{H_u}^2 - m_{H_d}^2 +
        2 ( m_{{Q_1}}^2  - m_{L_1}^2 - 2 m_{U_1}^2 +
	    m_{D_1}^2 + m_{E_1}^2 ) \nonumber \\ && \, + \, 
          ( m_{Q_3}^2 - m_{L_3}^2 - 2 m_{U_3}^2 
	  + m_{D_3}^2 + m_{E_3}^2 ) ,
\label{defS}
\end{eqnarray}
which vanishes at the GUT scale in the CMSSM,
is non-vanishing here and affects the running of most of the soft mass parameters below $\mgut$.
The behaviour of $m_{H_u}^2$ (green lines) and $m_{H_d}^2$ (red lines) in
Fig.~\ref{fig:RGEsCA10} is qualitatively similar in the two models, 
but weak-scale values are larger in magnitude in SU(5) due to the larger GUT-scale values. 
We emphasize that EWSB occurs not when $m_{H_u}^2$ turns negative, but rather when $m_{H_u}^2+\mu^2$ does,
because it is the latter combination that appears in the Higgs potential that develops a minimum 
(see {\it e.g.}~\cite{Baer:book,Drees:book}).

Sfermion masses are renormalized significantly more above $\mgut$, and
we notice big changes in the
behaviours of ${\stau_R}$ (magenta lines) and $\sel_R$ (tan lines). Their masses at
the electroweak scale are much larger in the model with $M_{in}= 2.4 \times 10^{18}$~GeV
than in the CMSSM. This reflects the large renormalization of the soft supersymmetry-breaking
masses of all the $\bf{10}$ sfermions by gauge interactions between $M_{in}$ and
$\mgut$. On the other hand, the $\stau_L$ (blue lines) and other $\bf{\overline 5}$ 
sfermions are slightly less renormalized between $M_{in}$ and $\mgut$ than the $\bf{10}$'s, 
so that the 
${\widetilde L}$ - ${\widetilde E}^c$ mass difference is smaller than in the CMSSM.
We also note the split of the values of the third-generation sfermion masses from 
those of the first two generations in the SU(5) model: 
it is most noticeable in the $\stau_R$ - $\sel_R$ mass difference. 
As mentioned in Ref.~\cite{Baer:2000gf}, 
this effect is due to the large Yukawa couplings of the third generation, 
and can be up to $\sim 20\%$. For
this reason, our analysis is not entirely equivalent to SU(5)-inspired studies 
(see {\it e.g.}  Ref.~\cite{Gogoladze:2008dk,su5inspired})
where $\mfiv,\ \mten$ ($=\mfivl,\ \mtenl$) 
and $m_{\calh_1,\calh_2}$ are treated as free parameters at $\mgut$.

The relatively large renormalization of the  $\stau_R$ mass
between $M_{in}$ and $\mgut$ has some important implications for the appearances
of the $(m_{1/2}, m_0)$ planes for large $M_{in}$. One, as we discuss in more detail below, 
is that the boundary where
the LSP becomes the $\stau_1$ recedes to lower $m_{1/2}$ and $m_0$, since
only for smaller values of these parameters is the super-GUT renormalization insufficient
to maintain $m_{\stau_1} > m_\chi$. As a consequence, the coannihilation strip also recedes to
lower $m_{1/2}$ and $m_0$, since it is only close to the $\stau_1$ LSP boundary
that $m_{\stau_1} - m_\chi$ is small enough for coannihilation to bring the relic density
down into the WMAP range~\footnote{In minimal SU(5) with unification, just as in the CMSSM, 
the $\stau_1$ is dominantly $\stau_R$. 
Coannihilation with $\stau_1 \simeq \stau_L$, as found in Ref.~\cite{Gogoladze:2008dk} in a 
SU(5)-inspired framework, requires $\mten/\mfiv \simeq 50$ at $\mgut$,
which cannot be realized in the unified framework studied here.}.

The bottom left panel of Fig.~\ref{fig:RGEsCA10} shows how the sparticle spectrum evolves as
a function of $M_{in}$ between $\mgut$ and $\overline{M_P}$, for the same
$\tan \beta = 10$, $m_{1/2} = 900$~GeV, $m_0 = 218$~GeV and $A_0 = 0$ as in the left
panel. Note the rapid monotonic increase in $m_{\stau_1} - m_\chi$ with $M_{in}$. Since
the change in $m_{\stau_R}^2$ is largely independent of $m_0$, depending only on
$m_{1/2}$, it is clear that, as $M_{in}$ increases, $m_{\stau_1} < m_\chi$ only for
ever-smaller values of $m_{1/2}$ and $m_0$. Likewise, only for ever-smaller values of 
$m_{1/2}$ and $m_0$ is $m_{\stau_1} - m_\chi$ small enough for coannihilation to
bring the relic density into the WMAP range.

\begin{figure}[ht]
\begin{center}
\epsfig{file=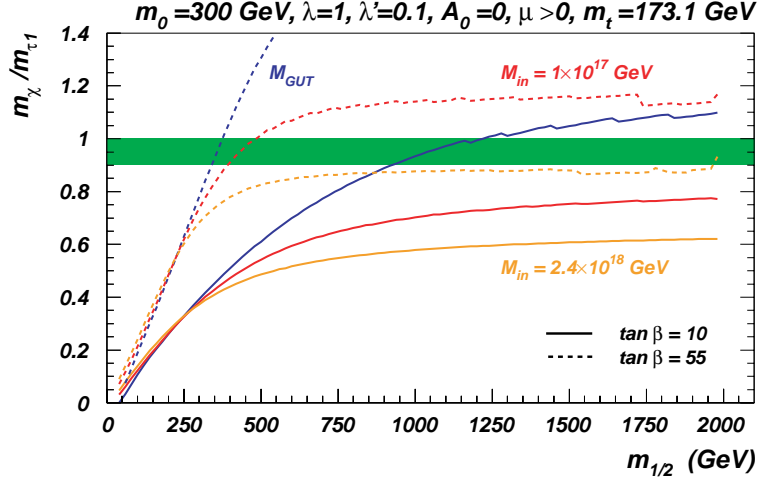,width=10cm}
\end{center}
\vskip -.2in
\caption{\it
The ratio neutralino to stau masses as a function of $m_{1/2}$ for three choices of $M_{in} = M_{GUT},
10^{17}$~GeV and $\overline{M_P}$ for the choices
$m_0 = 300$~GeV, $A_0 = 0$ and $\tan \beta = 10$ (solid) or 55 (dashed), 
assuming $\lambda = 1, \lambda' = 0.1$.  The shaded green horizontal band highlights the regime 
in which stau coannihilation is important.}
\label{fig:chi2stau}
\end{figure}

The effect of super-GUT renormalization of the masses of $\stau_1$ and $\chi$ is illustrated in 
more detail in Fig.~\ref{fig:chi2stau}, which shows the ratio $m_\chi/m_{\stau_1}$ as a 
function of $m_{1/2}$ 
for fixed $m_0 = 300$~GeV and different values of $M_{in}$. Whilst the ratio $m_\chi/m_{\stau_1}$ 
increases as a function of $m_{1/2}$ 
in each case, its slope with respect to $m_{1/2}$ is sensitive to both $M_{in}$ and $\tan \beta$.
For low values of $m_{1/2}$, the ratio is relatively independent of $M_{in}$ but sensitive to
$\tan \beta$.  However, for larger values of $m_{1/2}$  the super-GUT
renormalization depends strongly on $M_{in}$, and the $\stau_1$  receives a larger 
super-GUT contribution to its mass than does $\chi$, suppressing $m_\chi/m_{\stau_1}$. 
The green shaded horizontal strip highlights the area where
$0.9 \le m_\chi/m_{\stau_1} \le 1.0$, {\it i.e.}, the region where stau-coannihilation is expected to 
be important~\cite{stauco}. When $M_{in} = M_{GUT}$, $m_\chi/m_{\stau_1} > 1$ at $m_{1/2} = 1180$~GeV for
$\tan \beta = 10$, and coannihilation is important only for a small range of slightly
larger values of $M_{in}$. However, the ratio $m_\chi/m_{\stau_1} $
never exceeds 0.8 for $M_{in} > 10^{17}$~GeV, and $\stau$ coannihilation is not
effective for any value of $m_{1/2}$.  A similar trend is seen for $\tan \beta = 55$, though larger
$M_{in}$ is needed before coannihilation ceases to be effective.

The bottom right panel of Fig.~\ref{fig:RGEsCA10} shows the sensitivities to the value of $\lambda$
of the results for the point $m_0 = 300$~GeV, $A_0 = 0$ and $\tan \beta = 10$.
We see that they are quite weakly sensitive for this point, that was chosen
in the coannihilation strip region. As we mentioned earlier, RGE evolution for this 
point is dominated by gauge-gaugino terms, so a large $\lambda$ can only affect Higgs sector soft masses
and consequently $m_A$ and $\mu$, which are not relevant for neutralino annihilation in this case.

\begin{figure}
\epsfig{file=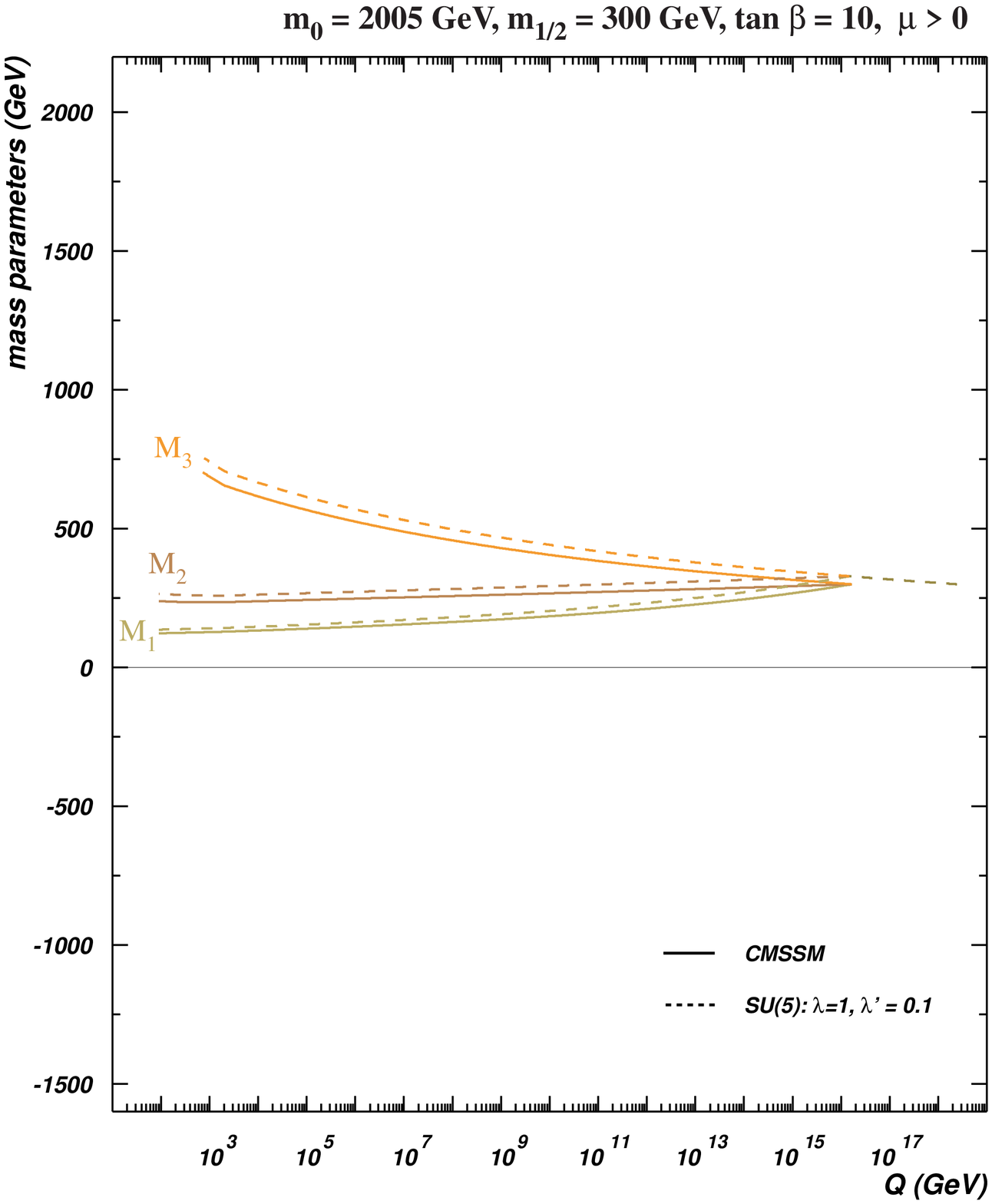,width=7cm}
\hskip .2in
\epsfig{file=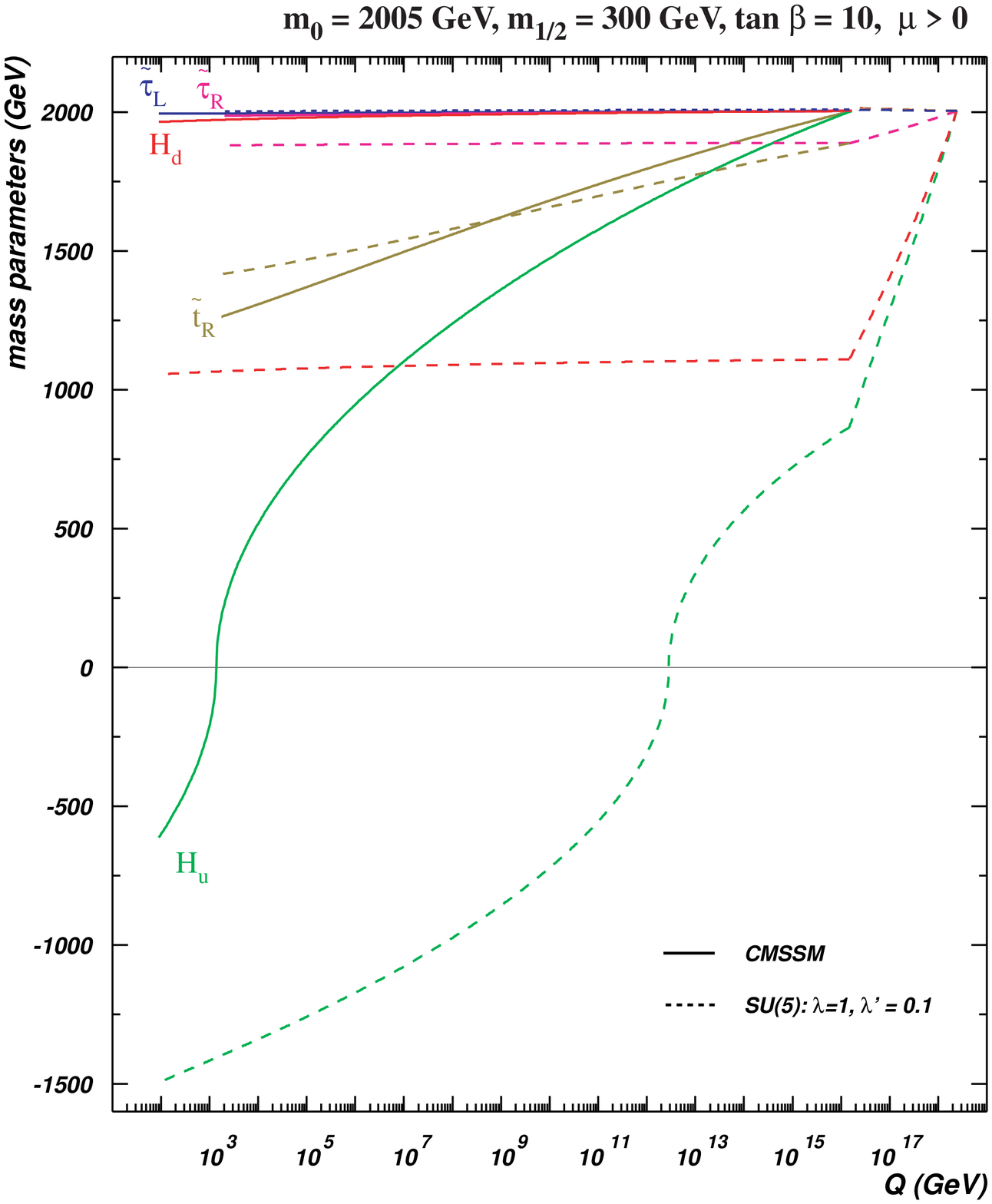,width=7cm} \\
\epsfig{file=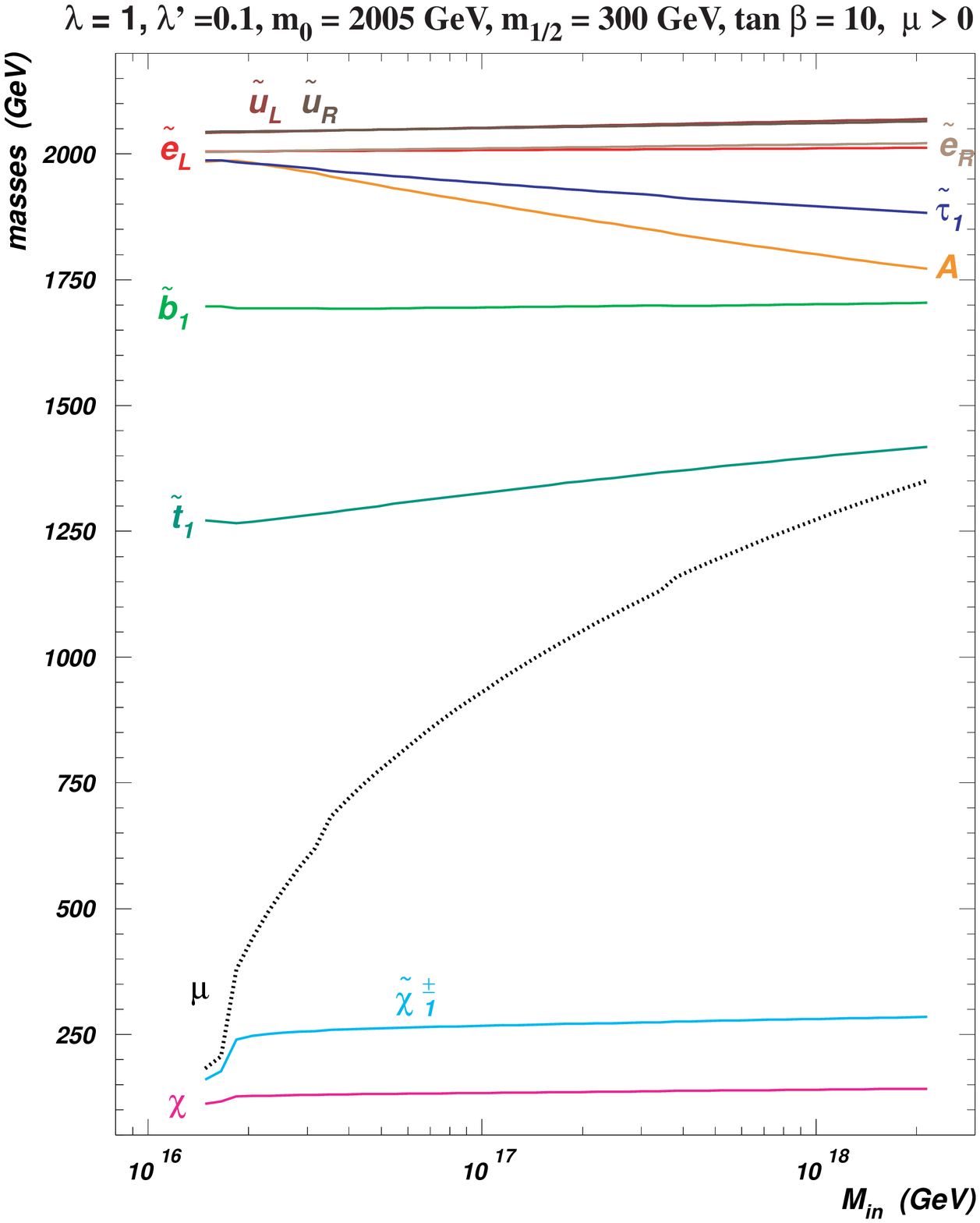,width=7cm}
\hskip .2in
\epsfig{file=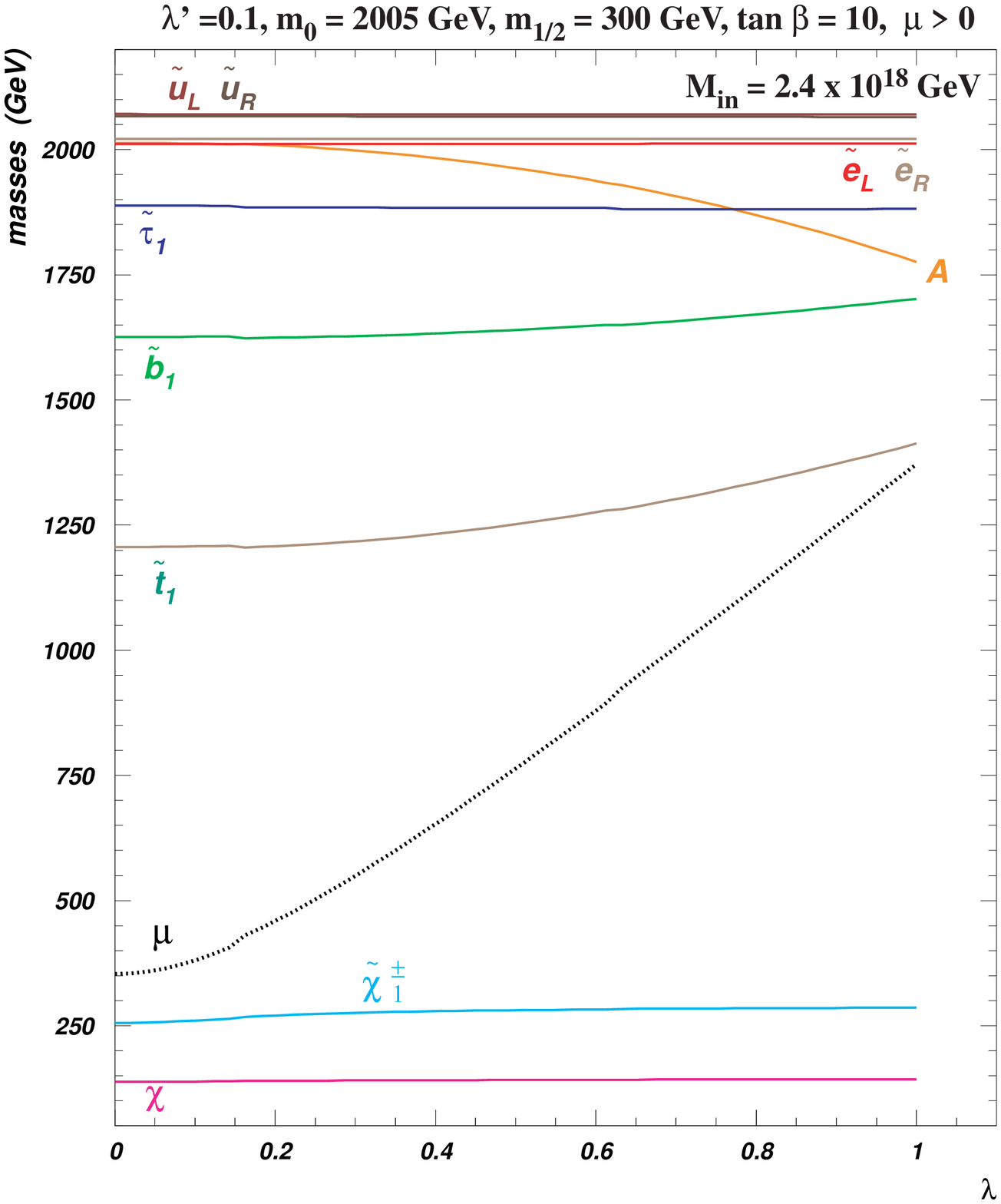,width=7.7cm}
\caption{\it
As for Fig.~\protect\ref{fig:RGEsCA10}, but for a focus point with
$m_0 = 2005$~GeV, $m_{1/2} = 300$~GeV, $A_0 = 0$ and $\tan \beta = 10$.}
\label{fig:RGEsFP10}
\end{figure}

Fig.~\ref{fig:RGEsFP10} shows a similar set of panels for a typical focus point with
$m_0 = 2005$~GeV, $m_{1/2} = 300$~GeV, $A_0 = 0$ and $\tan \beta = 10$ (equivalent to 
benchmark point E~\cite{Battaglia:2001zp}).
The upper right panel (for $\lambda = 1$) shows that in this case the most dramatic effects of increasing
$M_{in}$ are on $m_{H_u}^2$ (green lines) and $m_{H_d}^2$ (red lines). 
Since $m_{1/2} \ll m_0$, gauge-gaugino terms play only 
a subdominant role in the evolutions of these masses-squared.  
Therefore, if universality is assumed at $\mplr$ (dashed lines),
they are both renormalized significantly by $\lambda$ as $Q$ decreases from $\mplr$
to $\mgut$. Furthermore, at $M_{GUT}$ we have $m_{H_u}^2 < m_{H_d}^2$
and hence $S \simeq m_{H_u}^2 - m_{H_d}^2 < 0$ (neglecting here 
insignificant contributions from the squark mass splittings). 
(We recall that $S$ (\ref{defS}) enters into the MSSM
RGEs for $m_{H_u}^2$ and $m_{H_d}^2$, but is equal to 0 in the CMSSM.)  The effect of $S<0$
on the running of the Higgs soft masses is to push $|m_{H_u}^2|$ to higher values,
resulting in the larger value of $\mu$ parameter~\footnote{As a additional consequence, 
the scale at which $m_{H_u}^2$ is driven negative is
much larger than in the standard CMSSM $M_{in} = \mgut$ case (solid lines).}. This
increase in the value of $\mu$
suggests that the boundary of electroweak symmetry breaking recedes to larger $m_0$
for large $\lambda$, as we confirm later. 
We also see that the third-generation tenplet mass decreases significantly,
due to the effect of the large $\hten^2$ term.

The bottom left panel of Fig.~\ref{fig:RGEsFP10} shows that the low-scale sparticle
masses are relatively insensitive to $M_{in}$, though $m_{\stau_1}, m_A$
and $m_{\stop_1}$ do indicate some sensitivity. As noted above, there is a rapid
increase in $\mu$ as $M_{in}$ increases above $\mgut$. In the focus-point regime~\cite{focus}, 
the LSP has significant higgsino component so its mass $\simeq \mu$. 
As $M_{in}$ is increased, the LSP mass is once again $\simeq M_1$ and 
this implies that the relic density is larger than that allowed by
astrophysics and cosmology if $m_0$ is kept fixed.

The bottom right panel of Fig.~\ref{fig:RGEsFP10} shows how these effects depend
on the choice of $\lambda$. We note in particular that $\mu$ increases
significantly even for $\lambda \sim 0.3$. This suggests that the effects of
$\lambda$ on the locations of the focus-point region and the boundary of electroweak
symmetry breaking sets in already for moderate values of $\lambda$. 
We also see that the $\stau_1$ mass is insensitive to $\lambda$. 
Since $m_{\stau_R}^2$ is not significantly
renormalized below $\mgut$, we can confirm that the running of the
third-generation {\bf 10}-plet mass above
$\mgut$ is dominated by the large $\hten^2$ term. 
The observed sensitivities of the stop masses are entirely due to the
influence on MSSM running of changing boundary conditions at $\mgut$.

\begin{figure}
{\epsfig{file=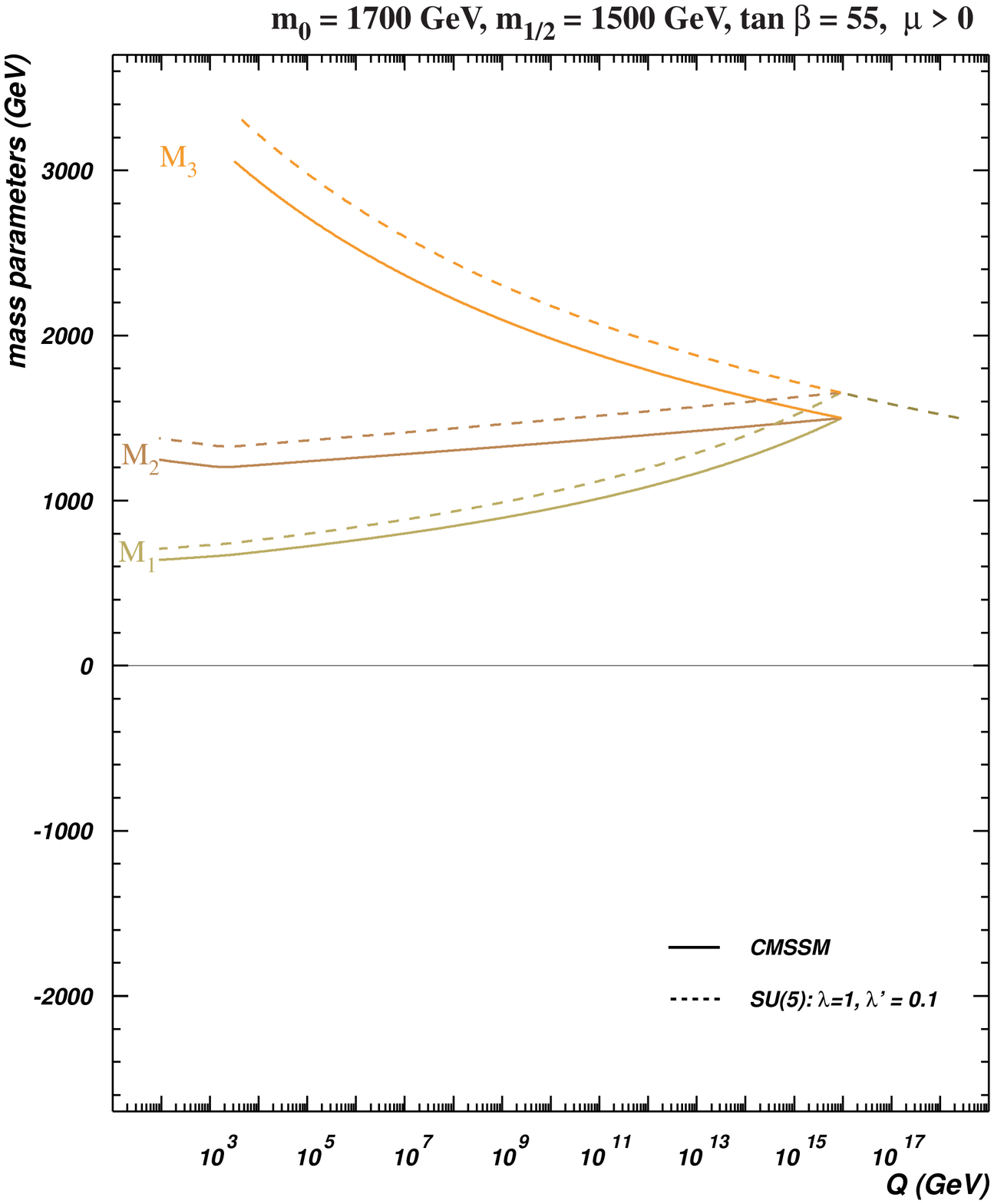,width=7cm}}
\hskip .2in
{\epsfig{file=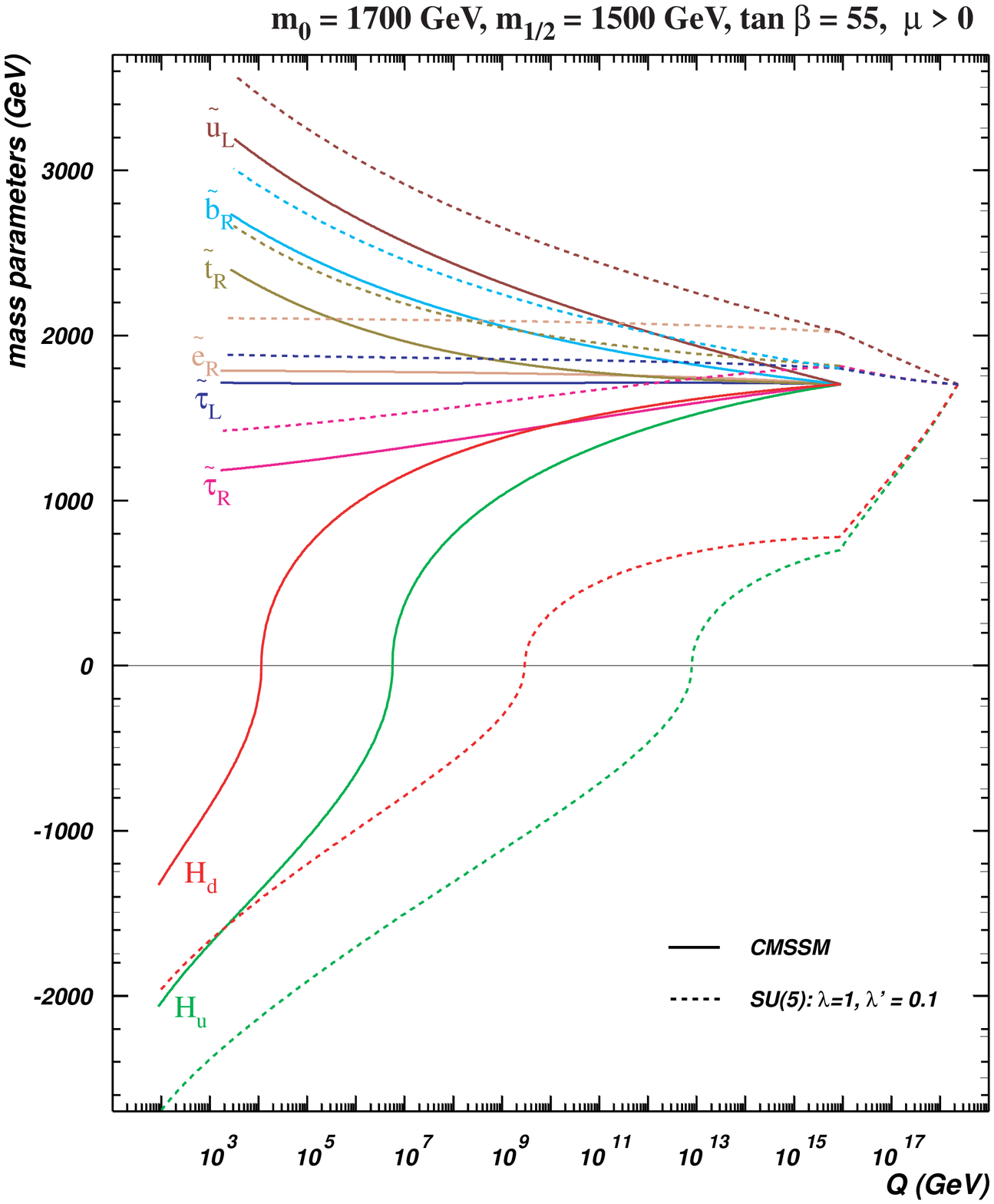,width=7cm}} \\
{\epsfig{file=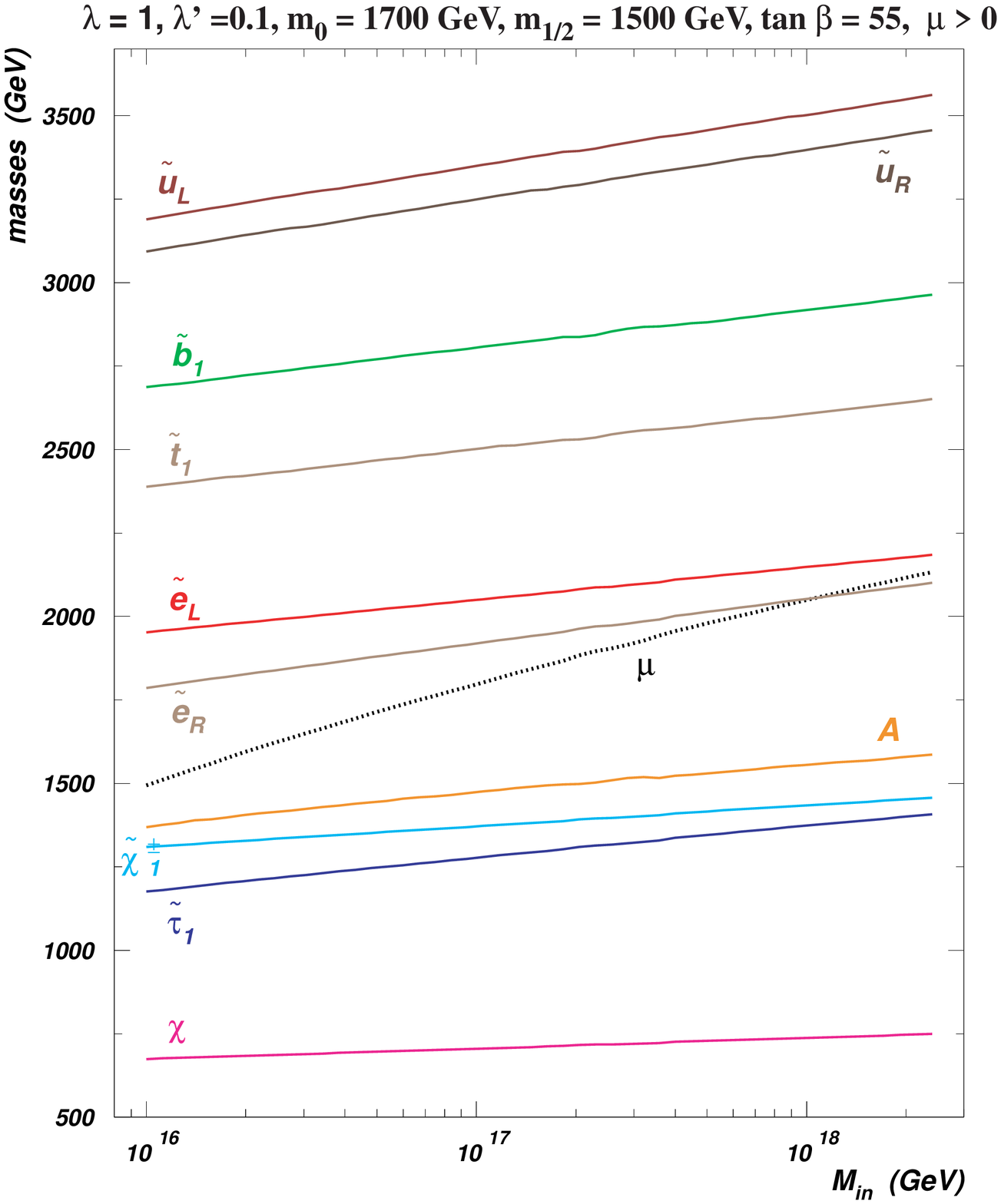,width=7cm}}
\hskip .2in
{\epsfig{file=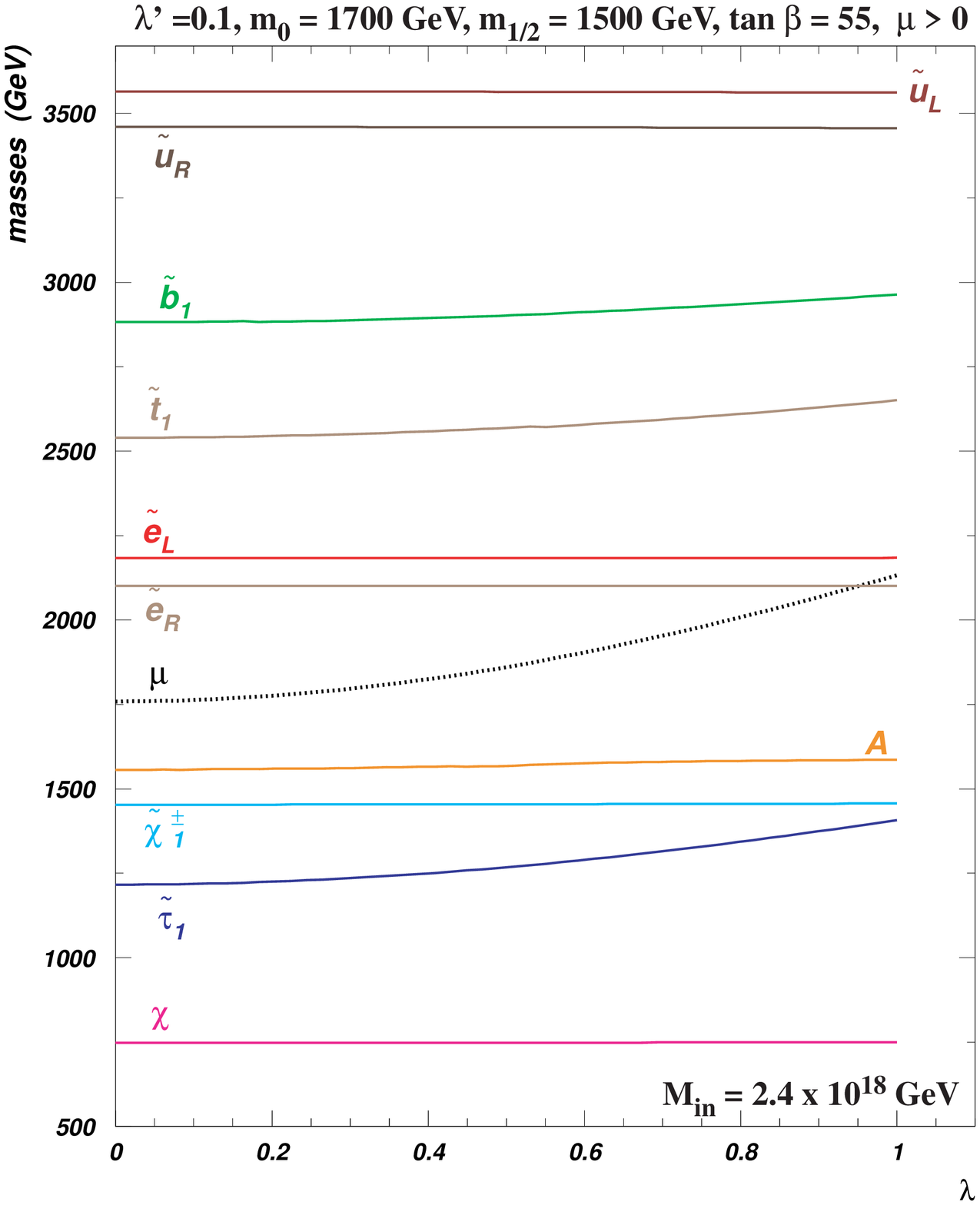,width=7.3cm}}
\caption{\it
As for Fig.~\protect\ref{fig:RGEsCA10}, but for a rapid-annihilation funnel point with
$m_0 = 1700$~GeV, $m_{1/2} = 1500$~GeV, $A_0 = 0$ and $\tan \beta = 55$.}
\label{fig:RGEsRA55}
\end{figure}

Fig.~\ref{fig:RGEsRA55} shows a similar analysis for a point in the rapid-annihilation 
funnel region~\cite{funnel} with $m_0 = 1700$~GeV, $m_{1/2} = 1500$~GeV, $A_0 = 0$ and $\tan \beta = 55$ 
(similar to benchmark point M~\cite{Battaglia:2001zp}).
We see in the top left panel that the effects of the renormalization between $\overline{M_P}$
and $\mgut$ are  important for $m_{H_u}^2$ (green lines) and
$m_{H_d}^2$ (red lines), and somewhat smaller for most sparticle masses.
In this region, the relic density is mainly controlled by the relation between
$m_A$ and $m_\chi$. We see in the bottom right panel of Fig.~\ref{fig:RGEsRA55}
that these vary in similar ways for $\mgut < M_{in} < \overline{M_P}$,
suggesting that the location of the rapid-annihilation funnel may be relatively
insensitive to $M_{in}$. Finally, we see in the bottom right panel of Fig.~\ref{fig:RGEsRA55}
that most of the sparticle masses are relatively insensitive to $\lambda$. The
exception is $m_{\stau_1}$, which is, however, of little relevance to the
relic density in this region of the $(m_{1/2}, m_0)$ plane. We see again that
the relation between $m_A$ and $m_\chi$ is quite stable, exhibiting little
sensitivity to $\lambda$.

\section{Representative \texorpdfstring{$(m_{1/2}, m_0)$}{(m12m0)} Planes}
\label{sec:plane}

We now discuss the consequences of the additional RGE evolution between $M_{in}$
and $\mgut$ for the CMSSM parameter space for some representative
$(m_{1/2}, m_0)$ planes for fixed $\tan \beta$ and $A_0$. In order to span the range
of plausible values of $\tan \beta$, we display planes for $\tan \beta = 10$ and 55.
We consider values of $M_{in}$ up to 
$\mplr \equiv \mpl/\sqrt{2\pi} \sim 2.4 \times 10^{18}$~GeV,
but restrict our attention to $A_0 = 0$. We assume $m_t = 173.1$~GeV~\cite{mt} and
$m_b^{\overline{MS}}(m_b) = 4.2$~GeV~\cite{rpp}. In all these $(m_{1/2}, m_0)$ planes,
we indicate by brown shading the regions that are excluded because the LSP is the lighter 
$\tilde \tau$. Regions where consistent vacuum breaking electroweak symmetry does
not occur are indicated by orange shading. 
We compute $BR(b \to s \gamma)$ as in \cite{gam}, 
and treat errors according to the procedure outlined in Ref.~\cite{bsgprocedure} when
studying the compatibility with the experimentally measured value~\cite{hfag};
regions excluded at 95\%CL are shaded green.
Regions favoured by $g_\mu - 2$ measurements~\cite{newBNL,g-2} if the Standard Model contribution
is calculated using low-energy $e^+ e^-$ data~\cite{Davier} are shaded pink, with the
$\pm 1\mhyphen\sigma$ contours shown as black dashed lines and the $\pm 2\mhyphen\sigma$ contours shown
as solid black lines. We restrict our attention to $\mu>0$, 
as suggested by both $b \to s \gamma$ and $g_\mu - 2$. 
The LEP lower limit on the chargino mass~\cite{LEPsusy} is shown as a thick black dashed
line, and the LEP lower limit on $m_h$~\cite{LEPHiggs} is indicated as a red dash-dotted line which
shows the position of the $m_h$ = 114.4~GeV contour. The light Higgs mass is
computed using the {\tt FeynHiggs~2.6.5} code~\cite{FeynHiggs}, 
and recall that its nominal results
should be assigned a theoretical error $\sim 1.5$~GeV, so that the location of the constraint
contour is only approximate. Finally, we use blue colour to indicate the regions where the
neutralino relic density falls within the 2-$\sigma$ WMAP range~\cite{WMAP}, 
$0.097 \leq \Omega_{CDM}h^2 \leq 0.122$.

\begin{figure}
\epsfig{file=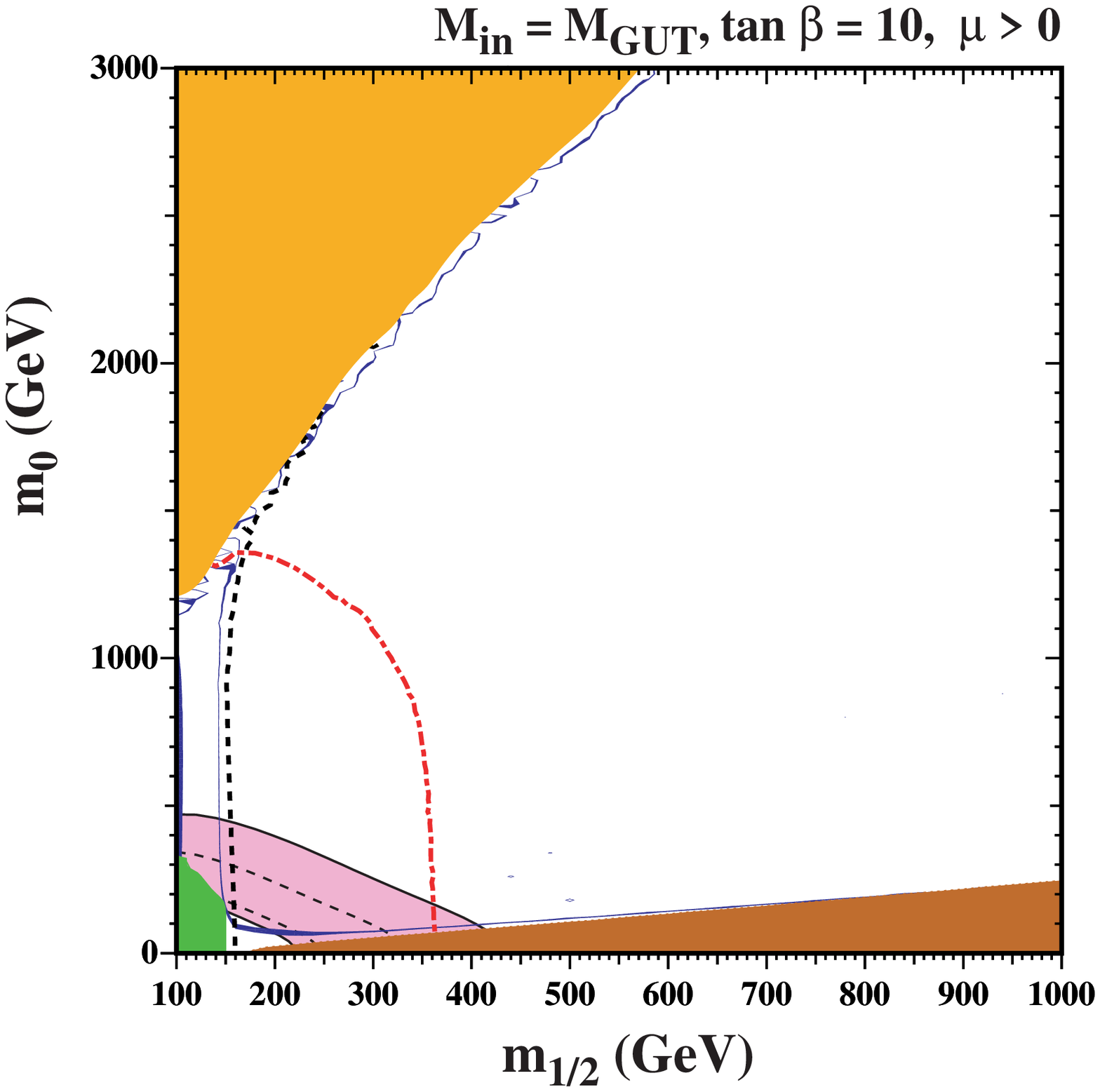,height=8.0cm}
\epsfig{file=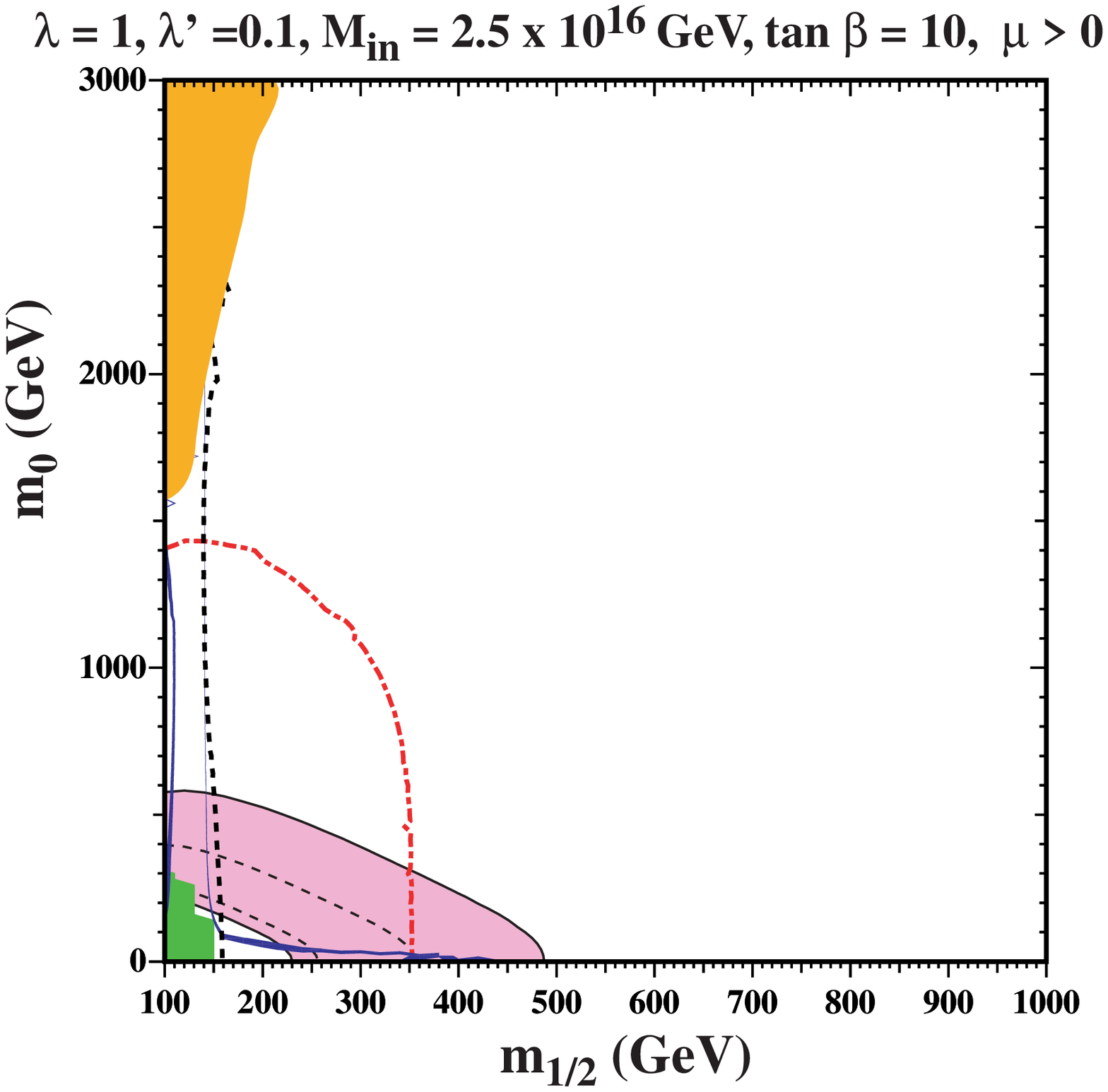,height=8.0cm}\\
\epsfig{file=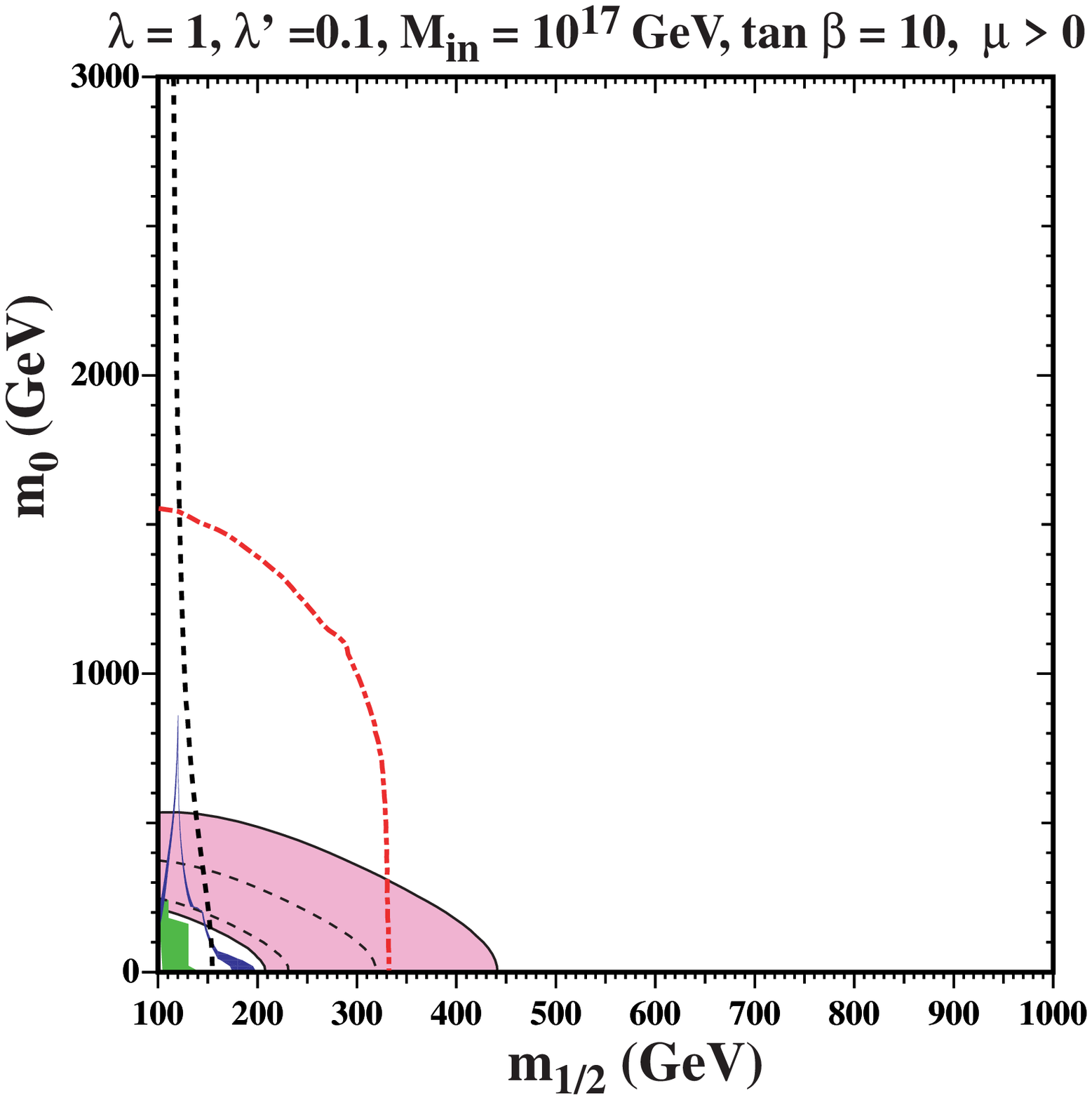,height=8.0cm}
\epsfig{file=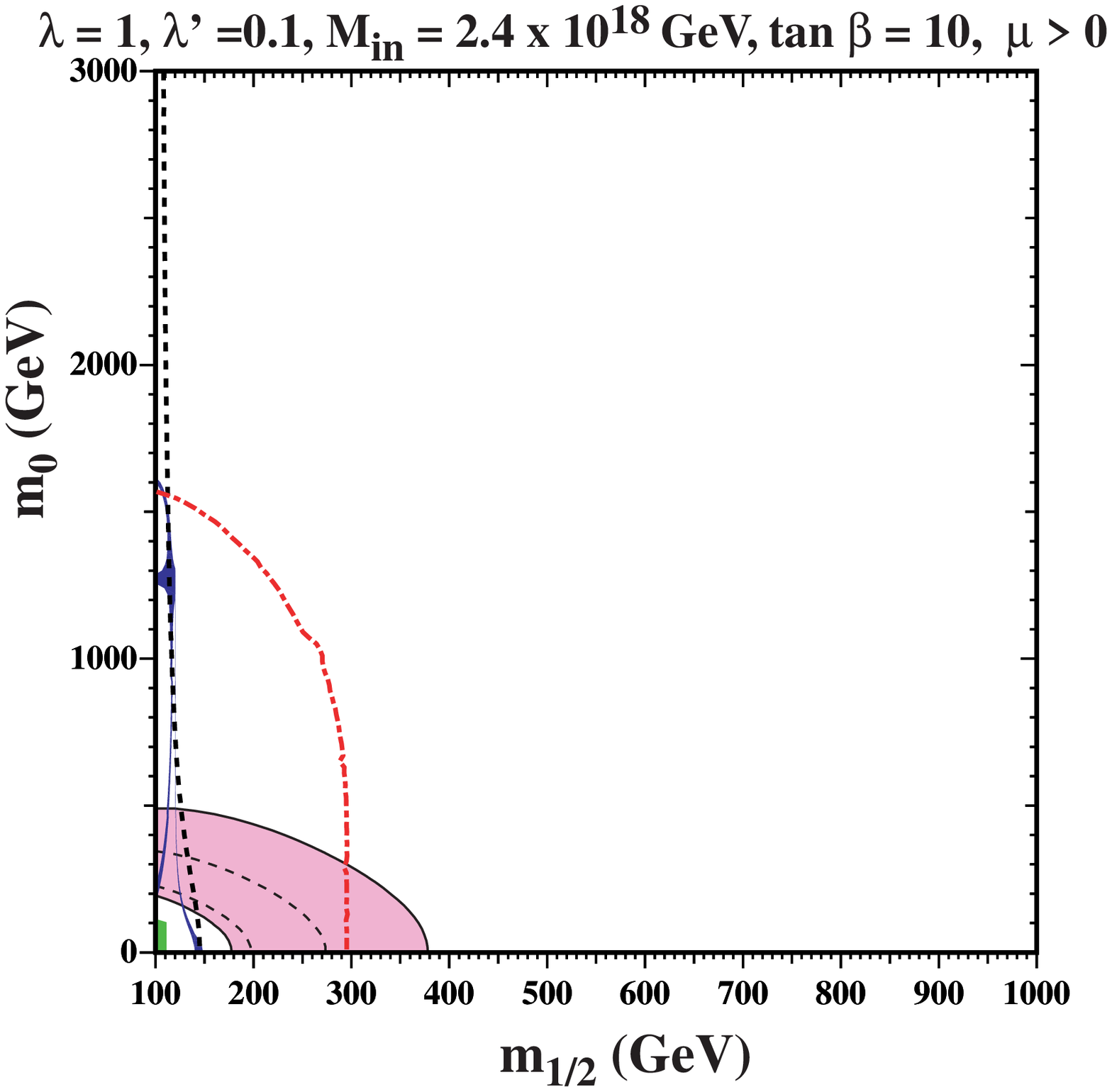,height=8.0cm}
\caption{\it
The $(m_{1/2}, m_0)$ planes for $A_0 = 0, \tan \beta = 10, \lambda = 1$
and $\lambda' = 0.1$ for different choices of $M_{in}$:
$\mgut$ (top left), $2.5 \times 10^{16}$~GeV (top right ), $10^{17}$~GeV (bottom left), 
and $2.4 \times 10^{18}$~GeV (bottom right). 
In the blue regions, $\ohsq$ is within the WMAP range. 
Pink region between black dashed (solid) lines are allowed by $g_\mu-2$ at 1-$\sigma$ (2-$\sigma$). 
Dark orange, brown and green colored regions are excluded.
Areas to the left of thick black dashed and red dash-dotted lines are ruled out by LEP searches 
for charginos and light MSSM Higgs respectively. More details can be found in the text.}
\label{fig:tb10}
\end{figure}

Fig.~\ref{fig:tb10} displays representative $(m_{1/2}, m_0)$ planes for $\tan \beta = 10$
and various values of $M_{in}$ and the specific choices $\lambda = 1, \lambda' = 0.1$.
The choice of a relatively large value of $\lambda$ highlights its potential importance 
(a smaller value $\lambda = 0.1$ is discussed later), and our results are quite
insensitive to the value of $\lambda'$. The choices of $\lambda$ and $\lambda'$
are irrelevant for the choice $M_{in} = \mgut$ 
shown in the top left panel of Fig.~\ref{fig:tb10}. 
Here, we see the familiar features of a stau coannihilation strip
extending up to $m_{1/2} \sim 900$~GeV, close to the boundary with the stau LSP region,
and a focus-point strip close to the boundary of electroweak symmetry breaking. There
is some tension between the $m_h$ and $g_\mu - 2$ constraints, but a region
of the coannihilation strip with $m_{1/2} \sim 400$~GeV would be consistent  with both
constraints.

Turning now to the choice $M_{in} = 2.5 \times 10^{16}$~GeV, shown in the top
right panel of Fig.~\ref{fig:tb10}, we see two dramatic effects from the modest
increase in $M_{in}$. One is the rapid disappearance of the stau LSP region, which
has retreated to $m_0^2 < 0$~\footnote{See~\cite{tachyon}
for a discussion of the circumstances under which this might be cosmologically acceptable.},
as a direct result of the renormalization effects seen in the left panel of Fig.~\ref{fig:RGEsCA10}.
A similar effect was seen in \cite{Calibbi,CEGLR}.
Because the ratio $m_\chi/\mstau$ is reduced as $M_{in}$ in increased, as shown in Fig.~\ref{fig:chi2stau},
there is a corresponding shift in the coannihilation strip to smaller $m_0$ and $m_{1/2}$.
In the particular example shown, the coannihilation strip extends to $m_{1/2} \sim 450$~GeV,
and there is a healthy portion compatible with the $g_\mu - 2$ constraint~\footnote{The
chargino, $m_h$, $g_\mu - 2$ and $b \to s \gamma$
constraints are relatively stable in the $(m_{1/2}, m_0)$ plane with respect to changes in $M_{in}$.}.
We note, in particular, that there is a region with $m_0 = 0$~\cite{nosc1,nosc2} which 
is compatible with all the constraints for
$360\, {\rm GeV} \lappeq m_{1/2} \lappeq 450$~GeV: however, its existence is very
sensitive to the choice of $M_{in}$. The other noticeable feature (also seen in \cite{Calibbi}) 
of the top
right panel of Fig.~\ref{fig:tb10} is the retreat of the electroweak symmetry breaking
constraint to smaller $m_{1/2}$ and larger $m_0$. As we discuss later, this effect
is quite sensitive to the value of $\lambda$, whereas the fate of the coannihilation region is
relatively insensitive to its value.

For the choice $M_{in} = 10^{17}$~GeV, shown in the bottom left panel of Fig.~\ref{fig:tb10},
these effects are more pronounced: both the coannihilation and the focus-point strips
have disappeared entirely. There is a small piece of the $(m_{1/2}, m_0)$ plane where
the relic density falls within the WMAP range, but this is incompatible with $m_h$ and
gives too large a value of $g_\mu - 2$. Finally, for the choice $M_{in} = 2.4 \times 10^{18}$~GeV, 
shown in the bottom right panel of Fig.~\ref{fig:tb10}, the small residual WMAP region falls
foul also of the chargino mass constraint.

\begin{figure}
\epsfig{file=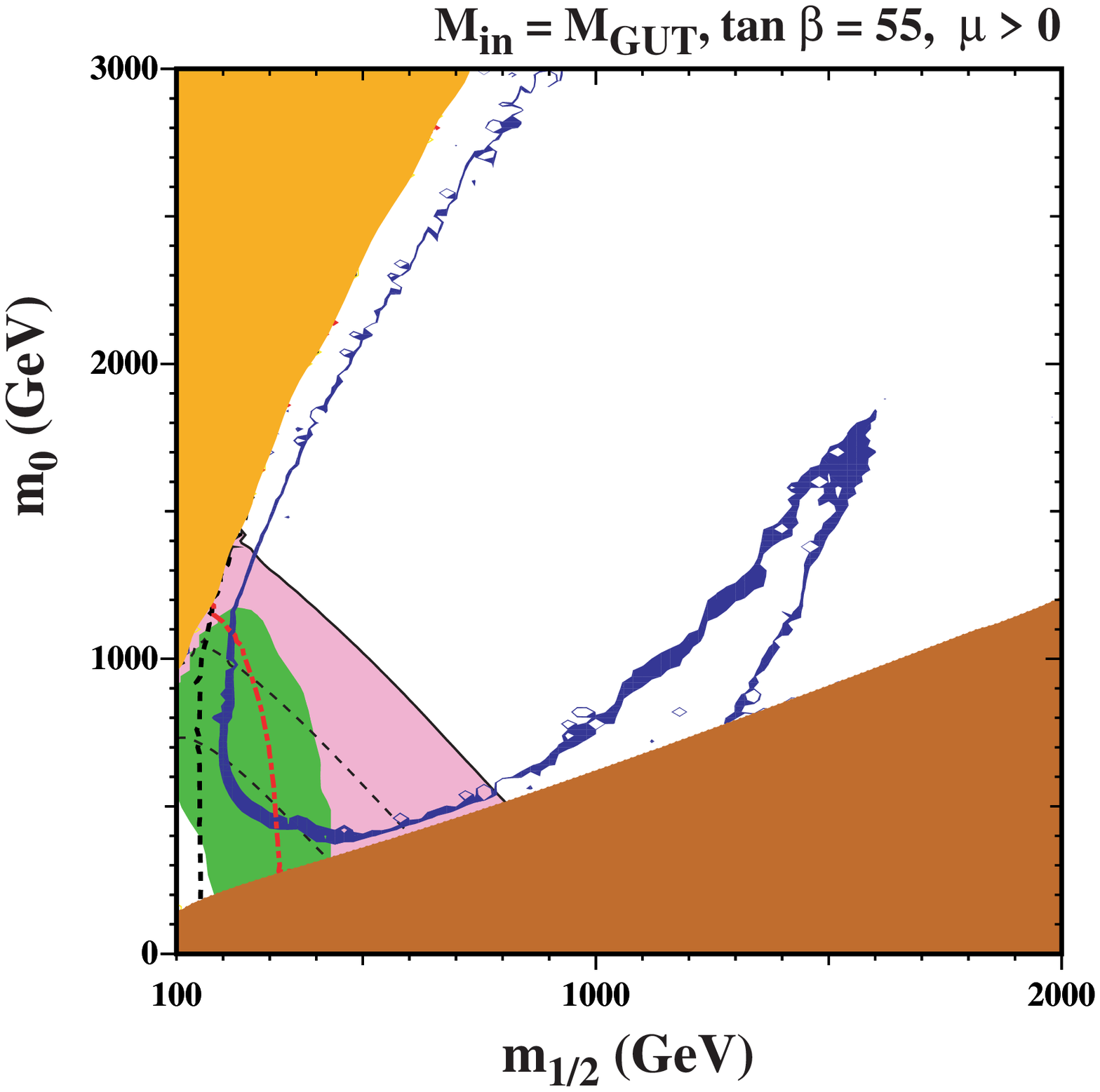,height=8.0cm}
\epsfig{file=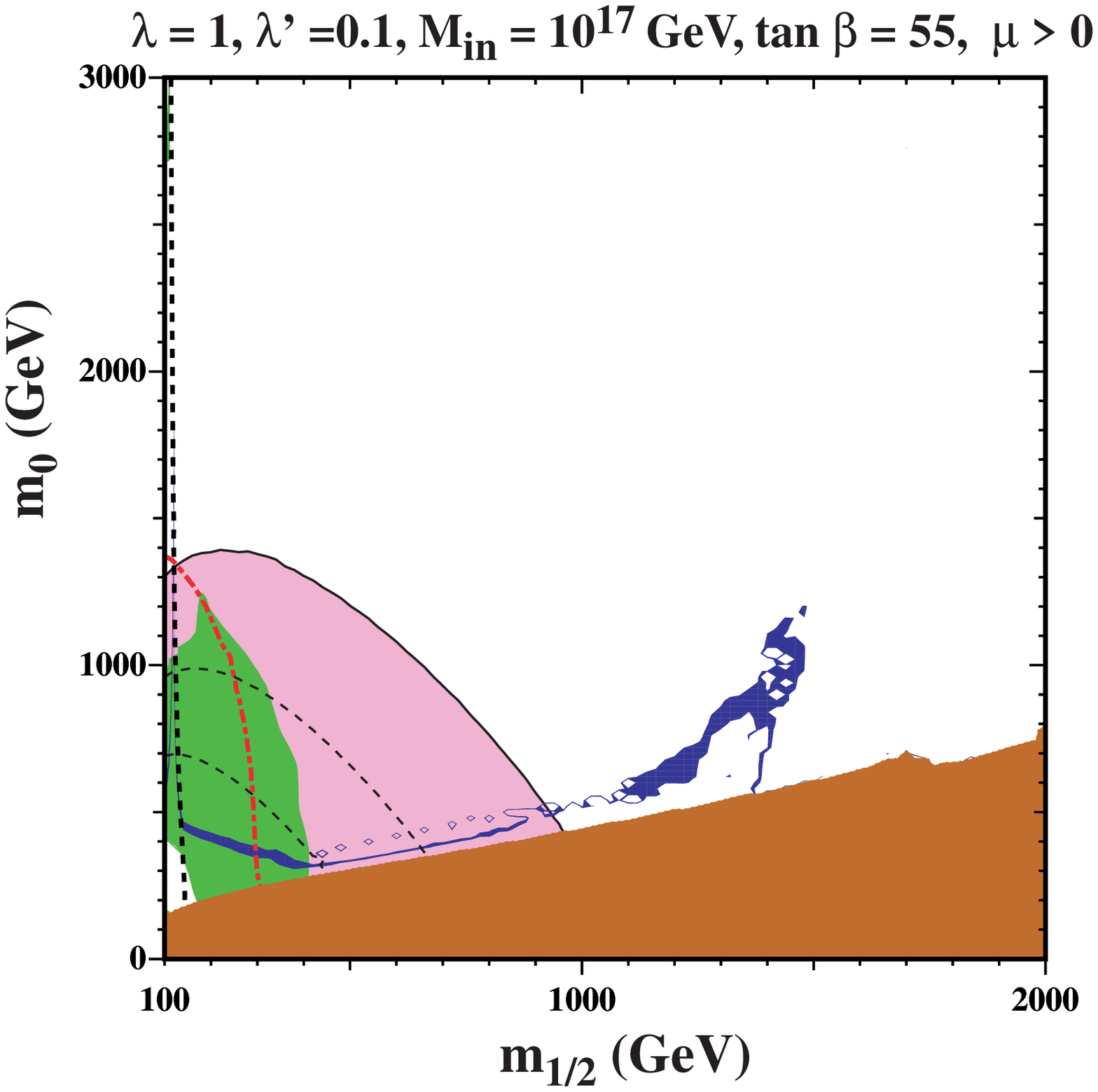,height=8.0cm}\\
\epsfig{file=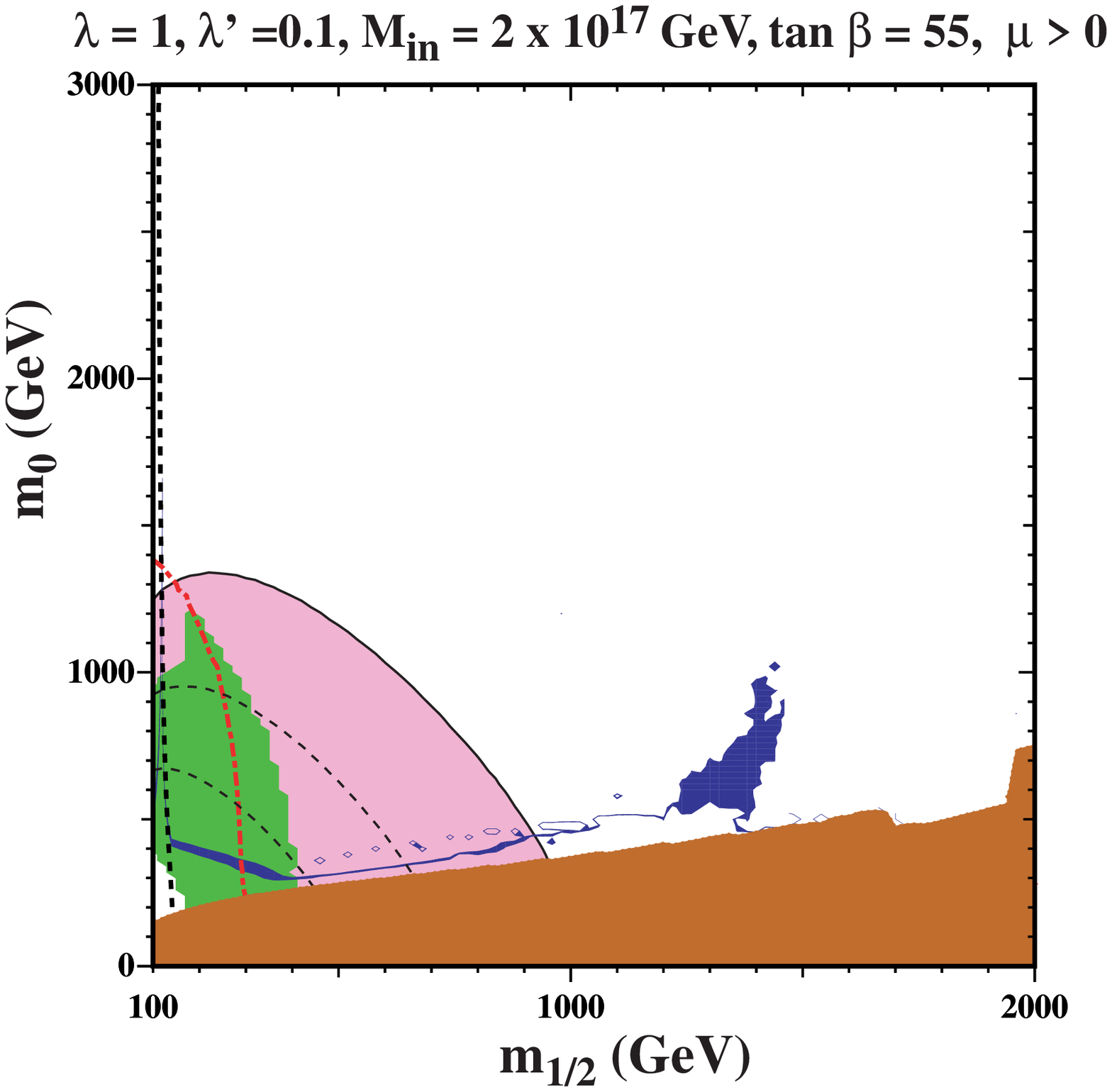,height=8.0cm}
\epsfig{file=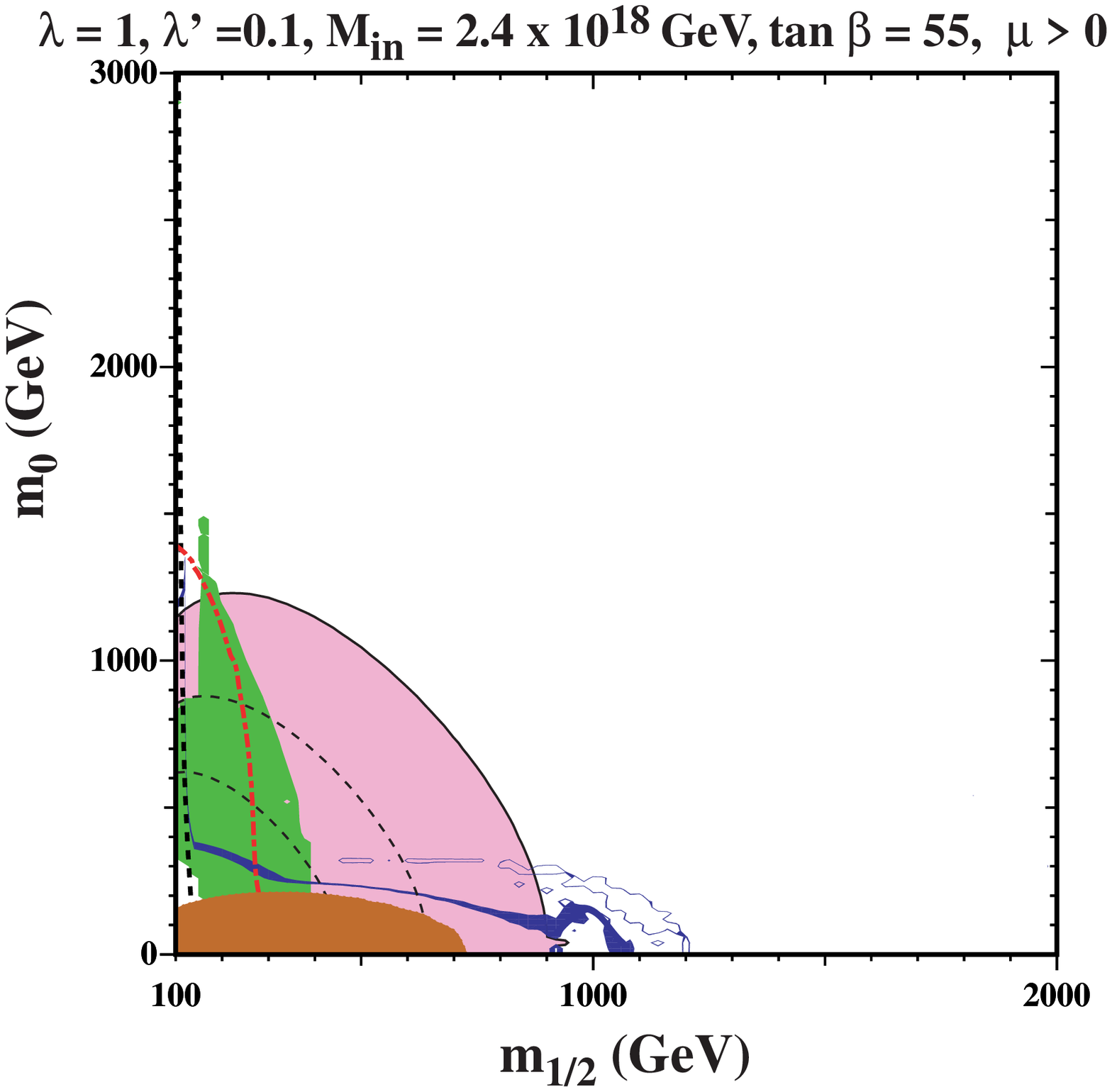,height=8.0cm}
\caption{\it
As for Fig.~\protect\ref{fig:tb10}, for $ \tan \beta = 55, \lambda = 1$
and $\lambda' = 0.1$ for different choices of $M_{in}$:
$\mgut$ (top left), $10^{17}$~GeV (top right), $2 \times 10^{17}$~GeV (bottom left), 
and $2.4 \times 10^{18}$~GeV (bottom right).}
\label{fig:tb55}
\end{figure}

We now consider the choice $\tan \beta = 55$, shown in Fig.~\ref{fig:tb55}. In this case,
in addition to the coannihilation and focus-point strips seen previously, we also note
the rapid-annihilation funnel that appears for $1000 {\rm GeV} \lappeq m_{1/2} \lappeq
1500$~GeV. As $M_{in}$ increases, the renormalization effects seen in the left panel
of Fig.~\ref{fig:RGEsCA10} cause the stau LSP region to retreat as in the $\tan \beta = 10$
case, though more slowly, and it does not disappear entirely, even for
$M_{in} = 2.4 \times 10^{18}$~GeV. Likewise, whilst the coannihilation strip shrinks
with increasing $M_{in}$, it does not disappear, and much of it remains consistent with
$m_h$, $b \to s \gamma$ and $g_\mu - 2$. The rapid-annihilation funnel also persists
as $M_{in}$ increases, staying in a similar range of $m_{1/2}$, but
shifting gradually to lower values of $m_0$. In particular, we note that for the case
$M_{in} =  2.4 \times 10^{18}$~GeV, the no-scale possibility 
$m_0 = 0$~\cite{nosc1,nosc2} is allowed, on one or both flanks of the rapid-annihilation funnel. 
Finally, we note that the
electroweak symmetry breaking boundary disappears entirely for the displayed choices of
$M_{in} > \mgut$, as does the focus-point WMAP strip.

\begin{figure}
\epsfig{file=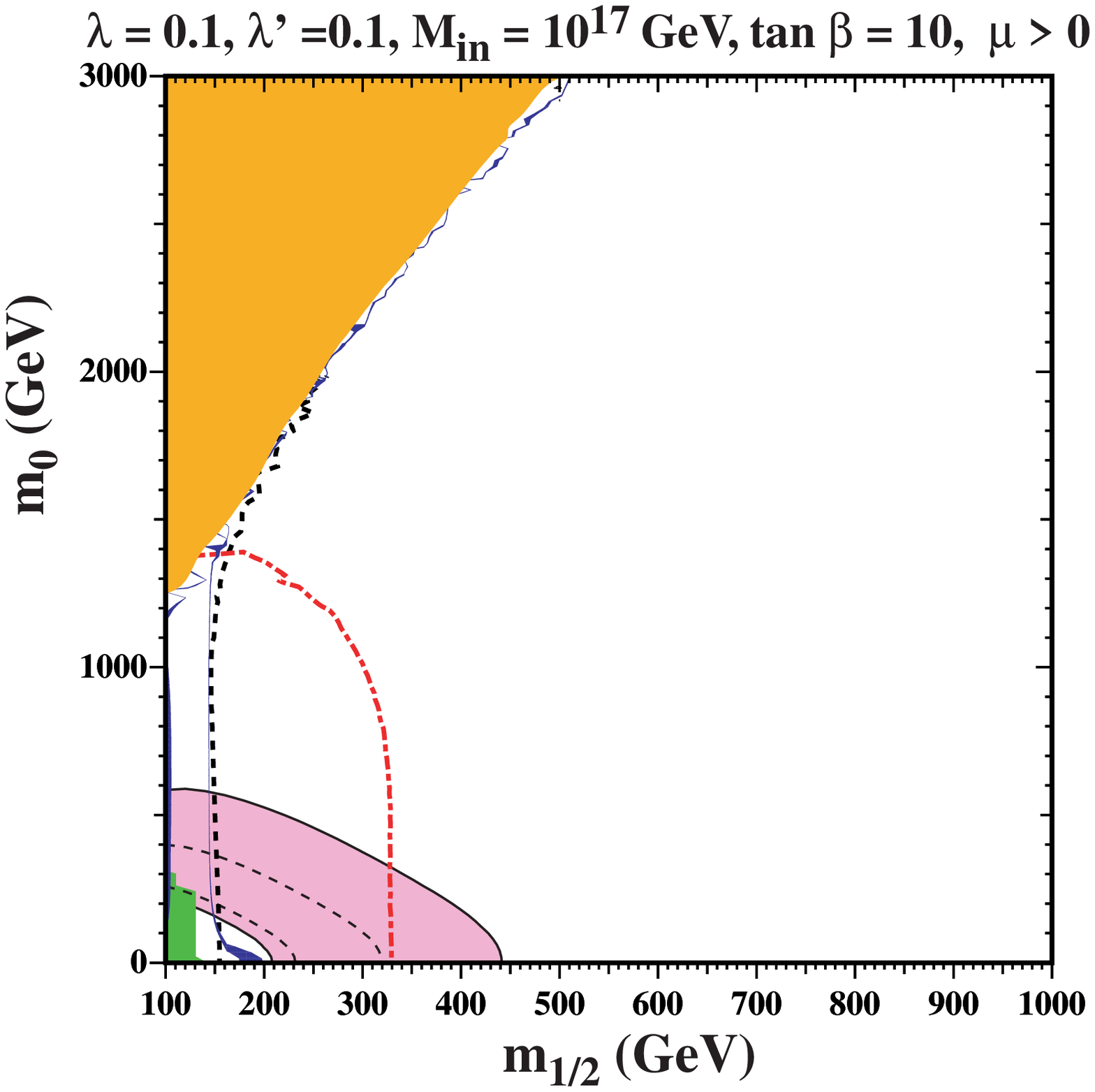,height=8.0cm}
\epsfig{file=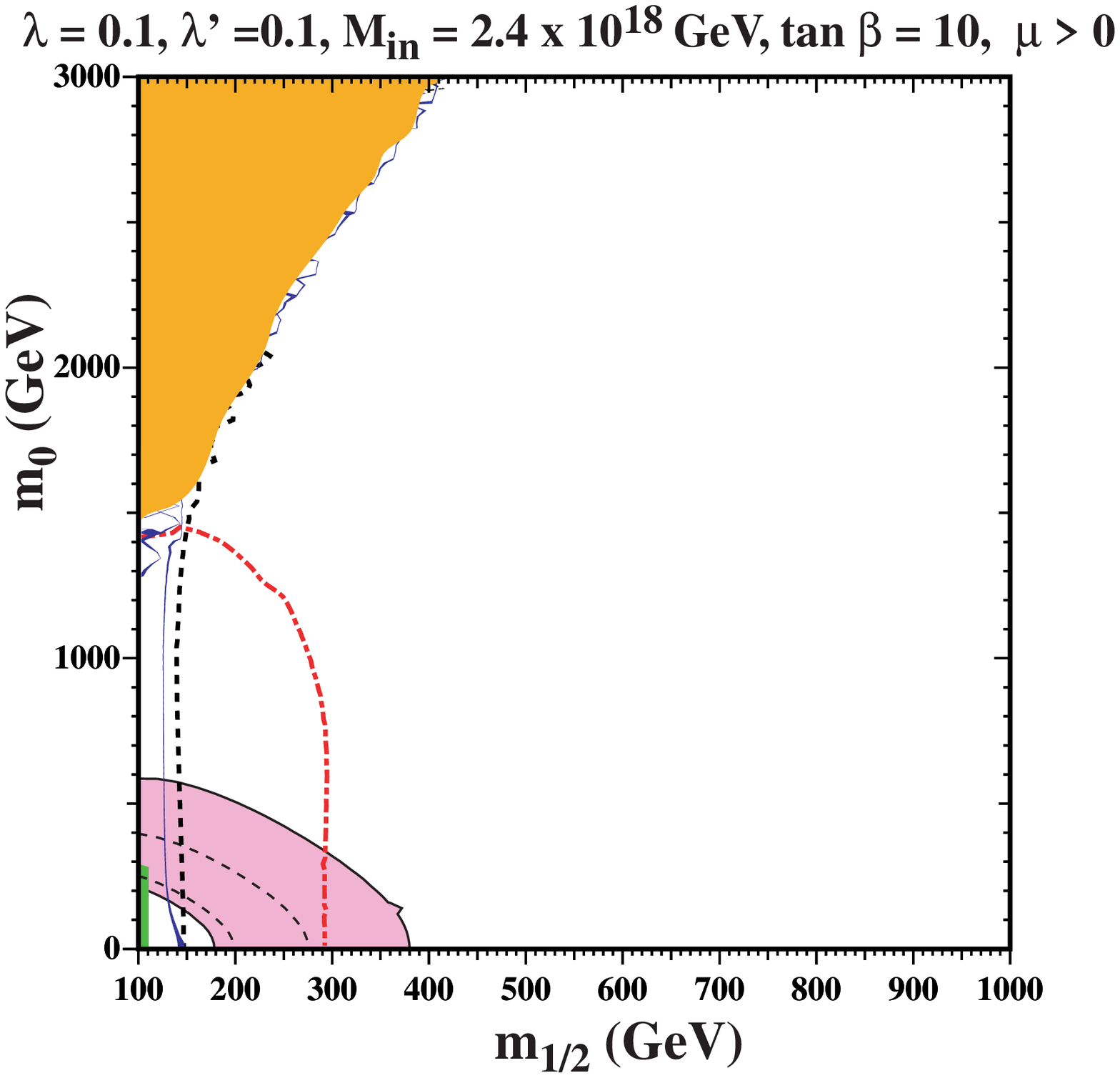,height=8.0cm}
\caption{\it
As for Fig.~\protect\ref{fig:tb10}, for $\tan \beta = 10, \lambda = 0.1$
and $\lambda' = 0.1$ for different choices of $M_{in}$:
$10^{17}$~GeV (left) and $2.4 \times 10^{18}$~GeV (right).}
\label{fig:tb10.1}
\end{figure}

In order to see the importance of the choice of $\lambda$, we display in
Fig.~\ref{fig:tb10.1} two examples of $(m_{1/2}, m_0)$ planes for the choice
$\lambda = 0.1$ and $\tan \beta = 10$, assuming $M_{in} = 10^{17}$~GeV 
and $2.4 \times 10^{18}$~GeV. Comparing with the bottom panels
of Fig.~\ref{fig:tb10}, we see very little change in the low-$m_0$ parts of the
planes: in particular, the stau LSP region and the coannihilation strip have
disappeared in the same ways. On the other hand, we see at large $m_0$
that the electroweak symmetry breaking boundary is very similar for the
two choices $M_{in} = 10^{17}$~GeV and $2.4 \times 10^{18}$~GeV, and that
these are in turn very similar to the corresponding boundary for $M_{in} = \mgut$,
shown in the top left panel of Fig.~\ref{fig:tb10}. This confirms that the the
differences in the focus-point regions shown in the other panels of Fig.~\ref{fig:tb10}
are due to the choice $\lambda = 1$ made there.

\begin{figure}[ht!]
\epsfig{file=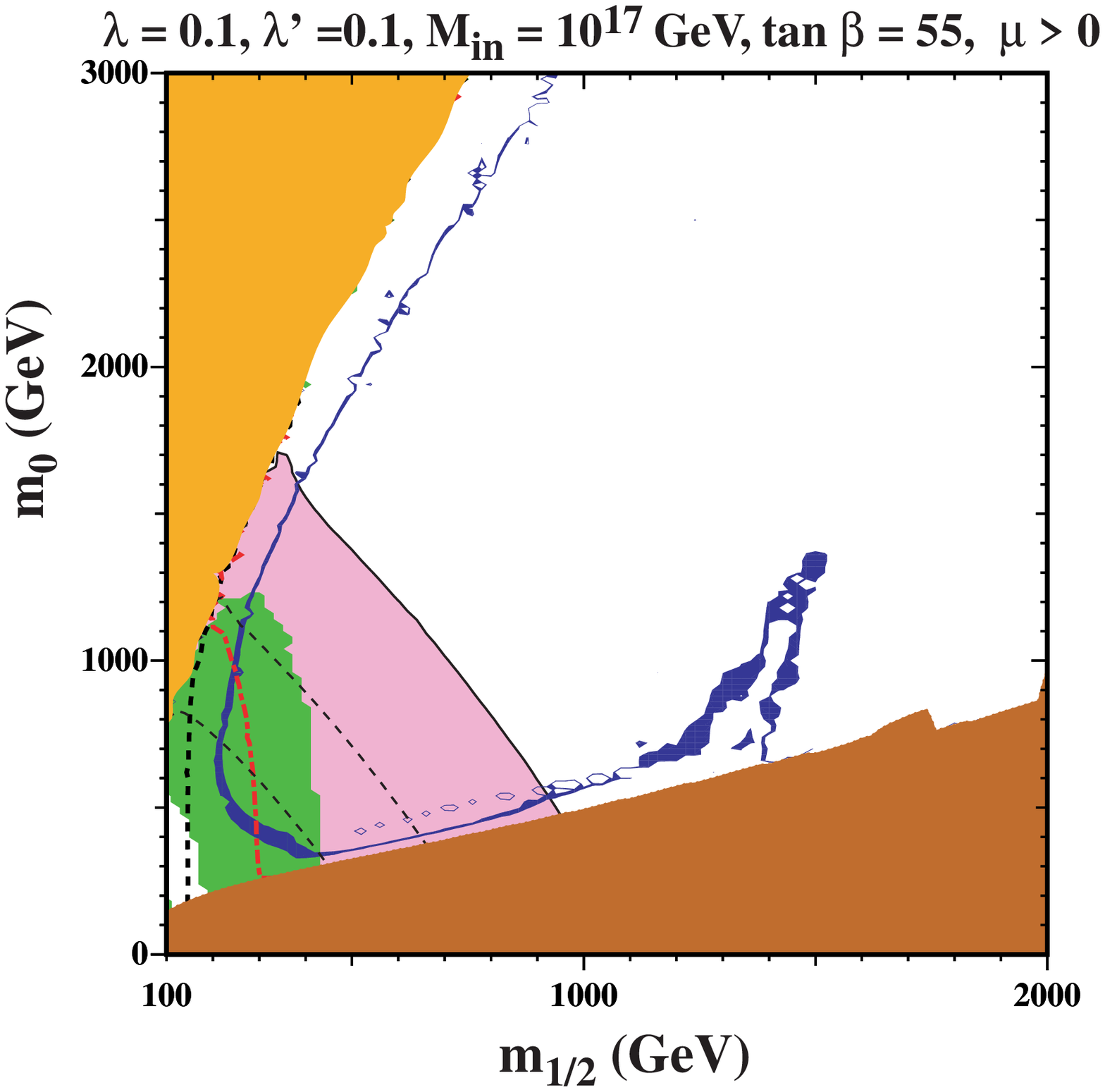,height=8.0cm}
\epsfig{file=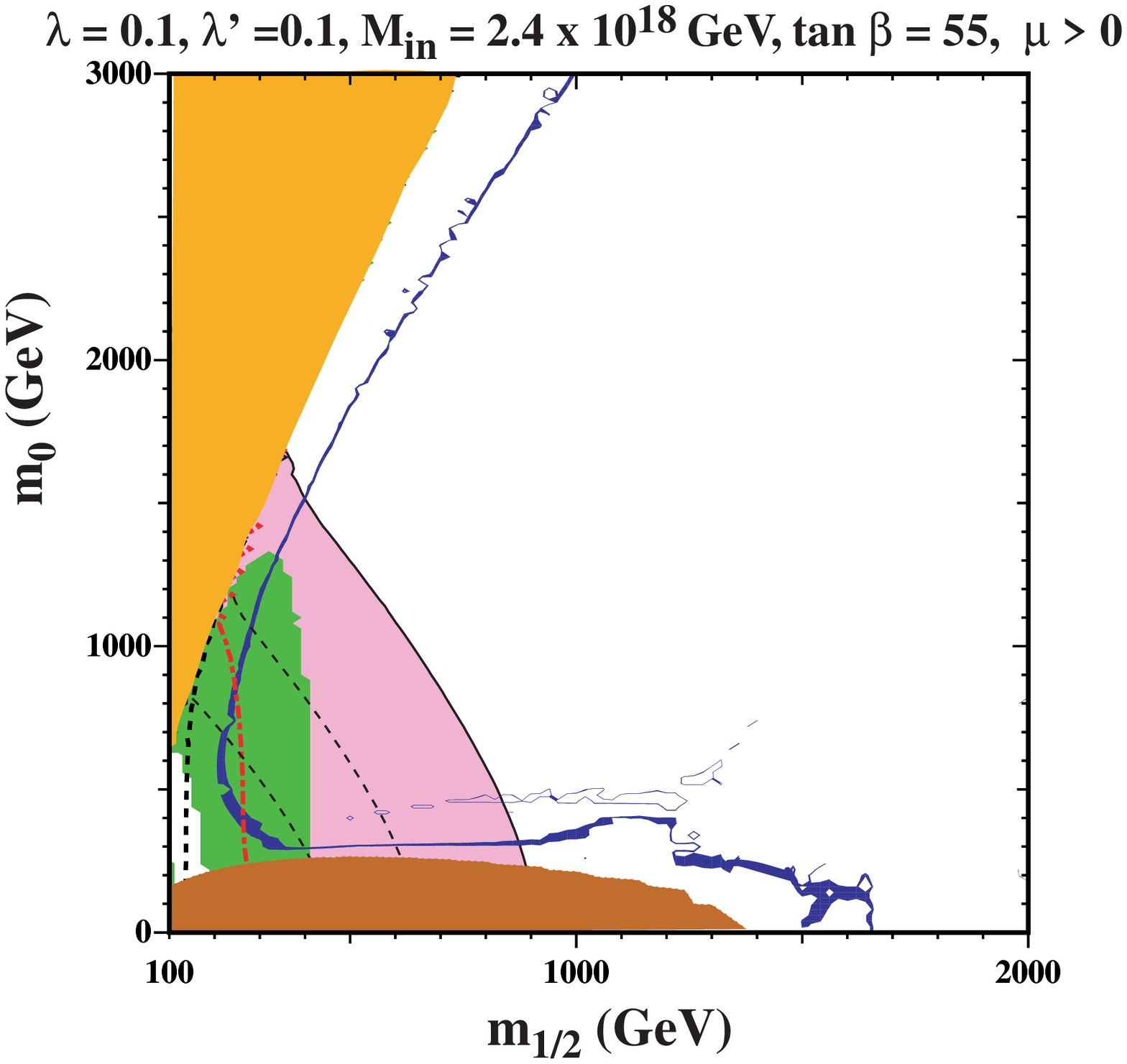,height=8.0cm}
\caption{\it
As for Fig.~\protect\ref{fig:tb10.1}, for $A_0 = 0, \tan \beta = 55, \lambda = 0.1$
and $\lambda' = 0.1$ for different choices of $M_{in}$:
$10^{17}$~GeV (left) and $2.4 \times 10^{18}$~GeV (right).}
\label{fig:tb55.1}
\end{figure}

Fig.~\ref{fig:tb55.1} shows a similar pair of comparisons for $\tan \beta = 55$, with
the choice $\lambda = 0.1$ and $M_{in} = 10^{17}$~GeV, $2.4 \times 10^{18}$~GeV.
Comparing with the corresponding (right) panels of Fig.~\ref{fig:tb55}, we see
that the stau LSP regions and the coannihilation strips are essentially identical,
indicating that the value of $\lambda$ is irrelevant in these regions, as expected.
However, differences again are found at large $m_0$, where the focus point 
region has made a reappearance where previously, it had
disappeared for all values of $M_{in} > \mgut$
when $\lambda = 1$, 
but is not only present for $\lambda = 0.1$, but has even moved to smaller values of $m_0$ than 
in the CMSSM case.

\begin{figure}
\epsfig{file=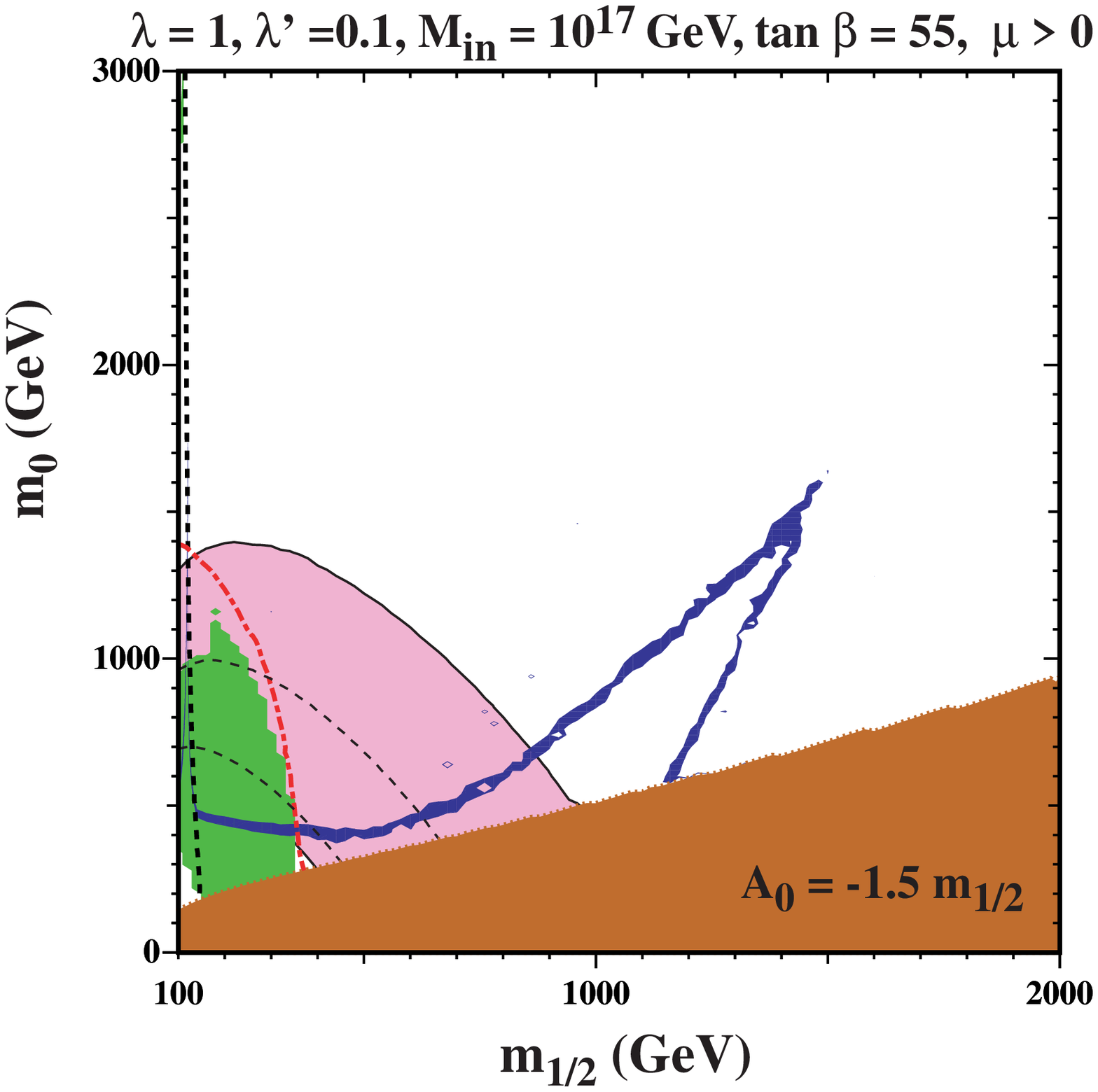,height=8.0cm}
\epsfig{file=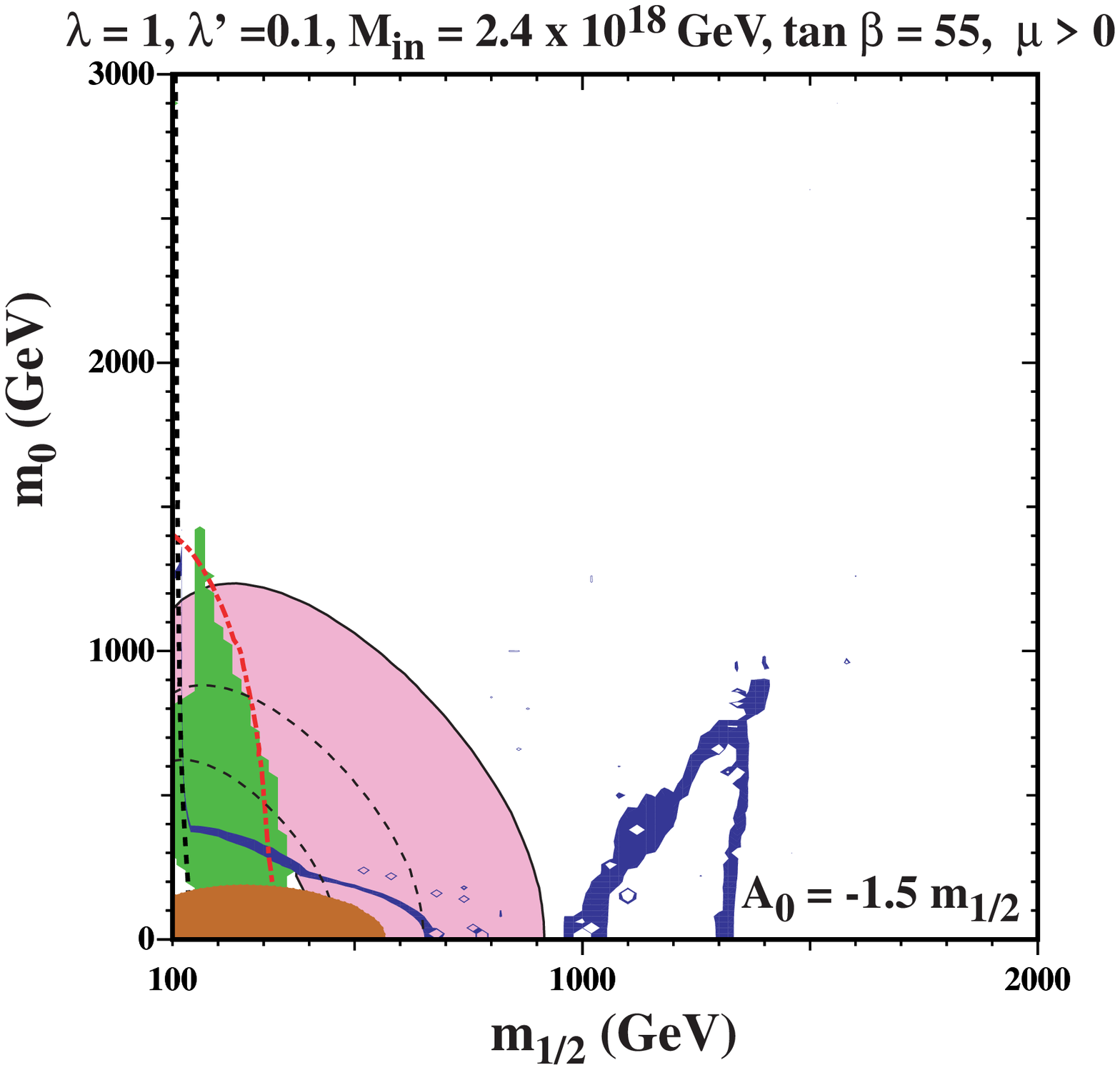,height=8.0cm}\\
\epsfig{file=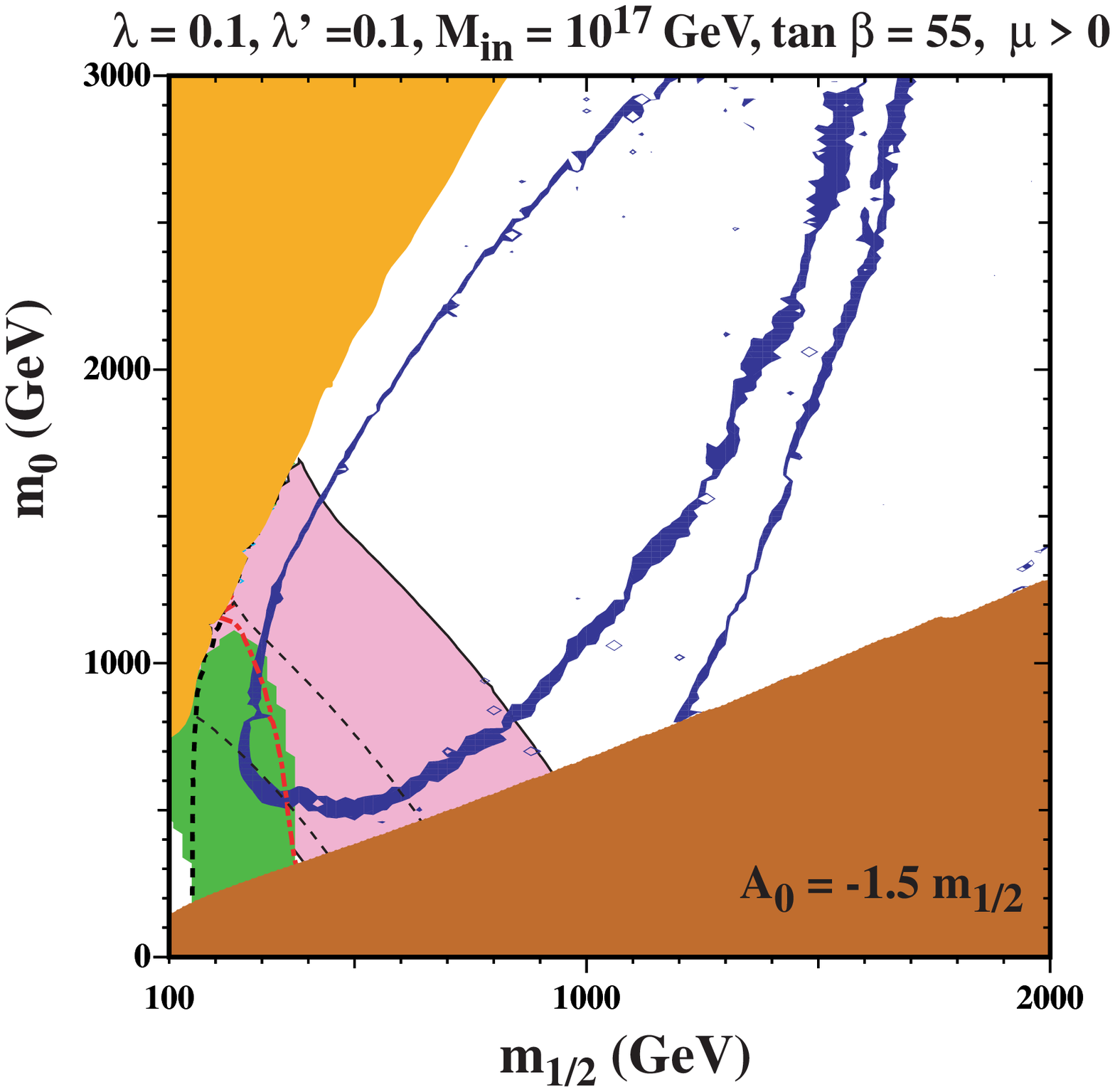,height=8.0cm}
\epsfig{file=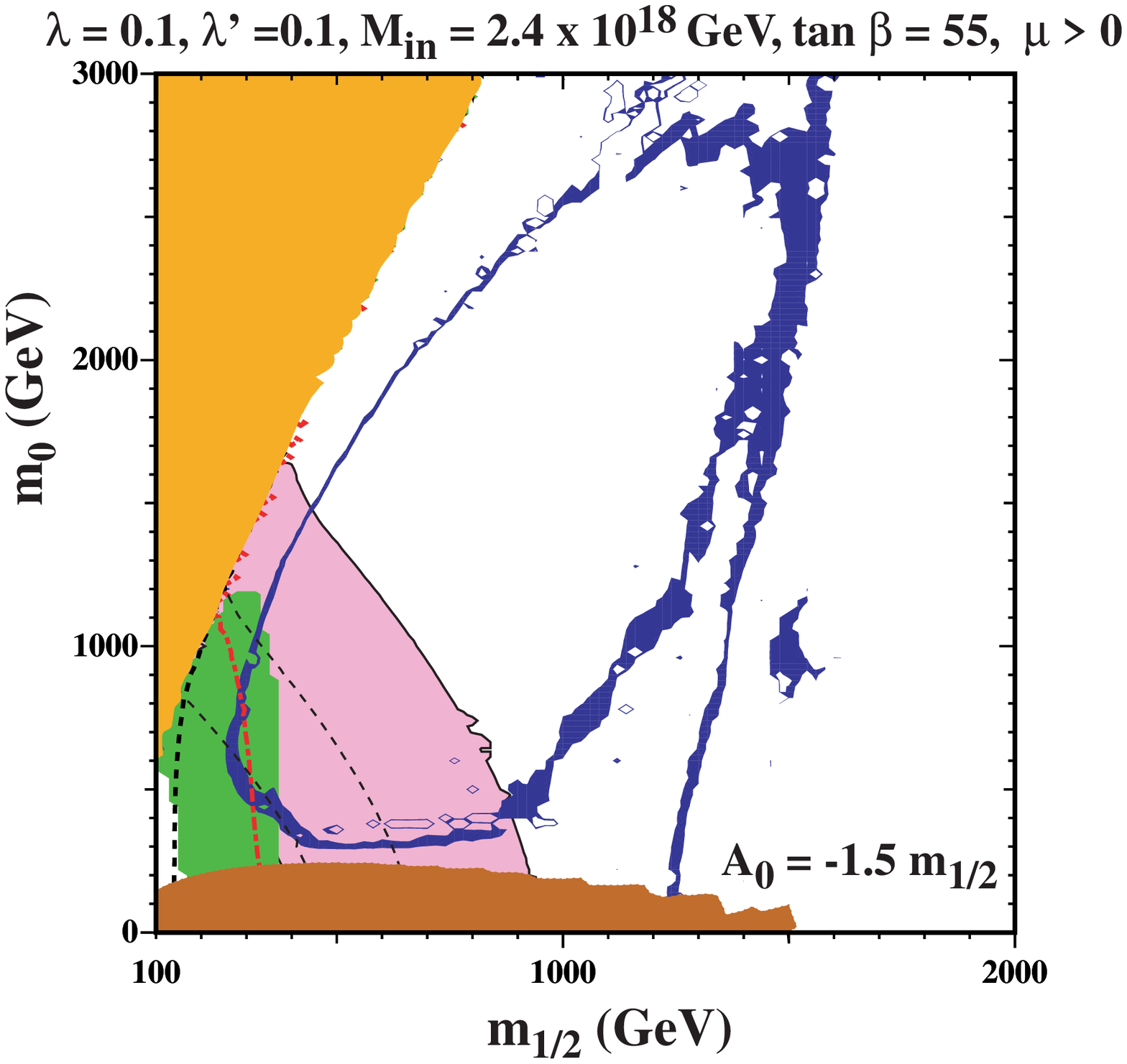,height=8.0cm}
\caption{\it
As for Fig.~\protect\ref{fig:tb10}, for $ \tan \beta = 55, A_0 = - 1.5 m_{1/2}, \lambda' = 0.1$
for $\lambda = 1$ (upper) and $\lambda = 0.1$ (lower) with $M_{in} = 10^{17}$~GeV (left), 
and $2.4 \times 10^{18}$~GeV (right).}
\label{fig:tb55ane0}
\end{figure}

The structure of the relic density regions when $\tan \beta = 55$ is quite sensitive to 
our assumption for $A_0$, as is also the case in the CMSSM, where
the funnel regions become more pronounced when $A_0 < 0$~\cite{ehow4}.
In Fig.~\ref{fig:tb55ane0}, we illustrate this effect by taking $A_0 = -1.5 m_{1/2}$ for 
 $M_{in} = 10^{17}$~GeV (left), $2.4 \times 10^{18}$~GeV (right)
and $\lambda = 1, 0.1$ (upper and lower, respectively).
In each case, when $A_0 < 0$
the funnel region reaches to higher values of $m_0$ and $m_{1/2}$
and its two branches are more clearly separated. When $\lambda = 0.1$, we
also see that the rapid-annihilation funnel region
begins to connect with the focus-point region. 

\begin{figure}[ht!]
\epsfig{file=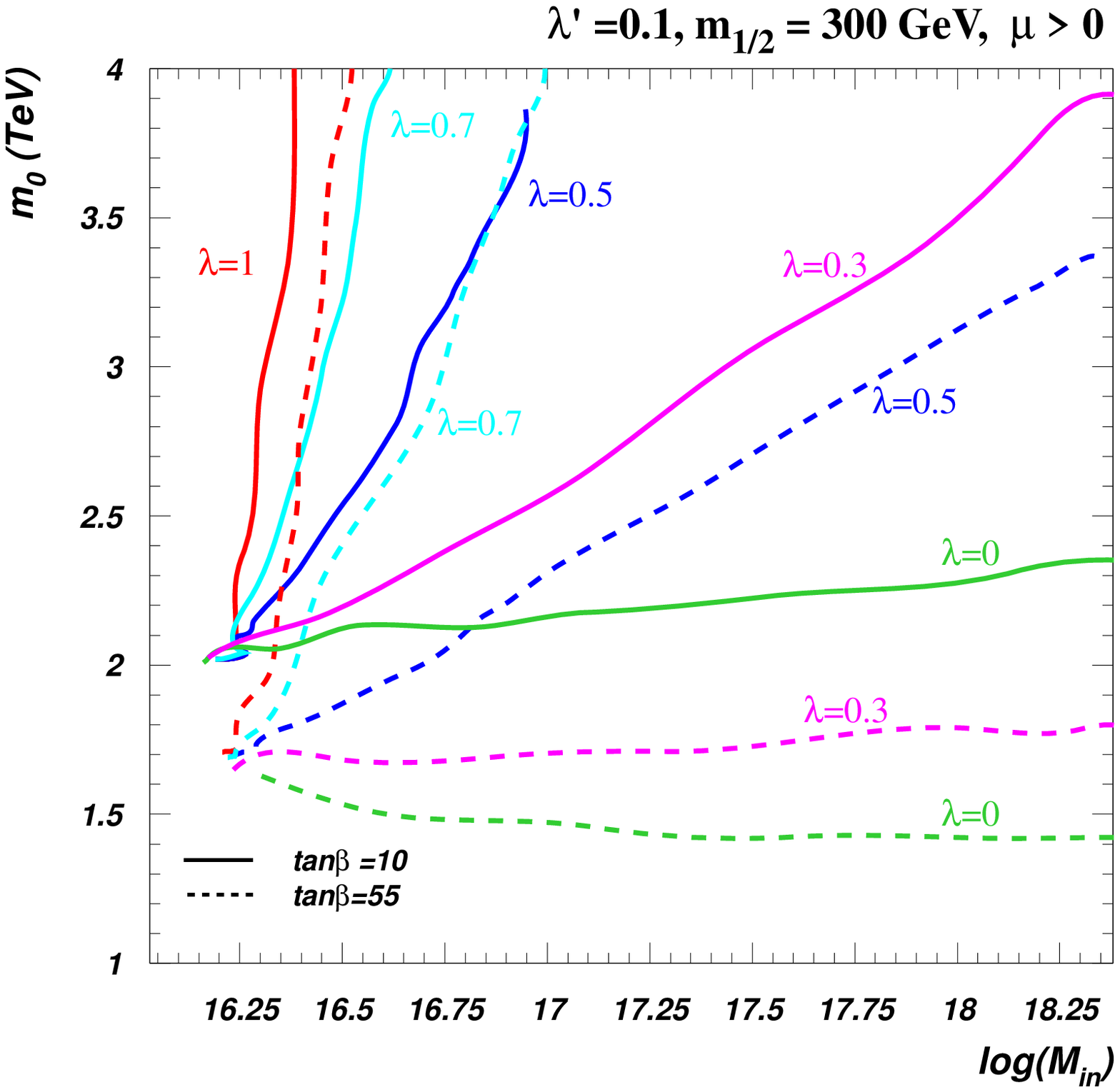,height=8.0cm}
\epsfig{file=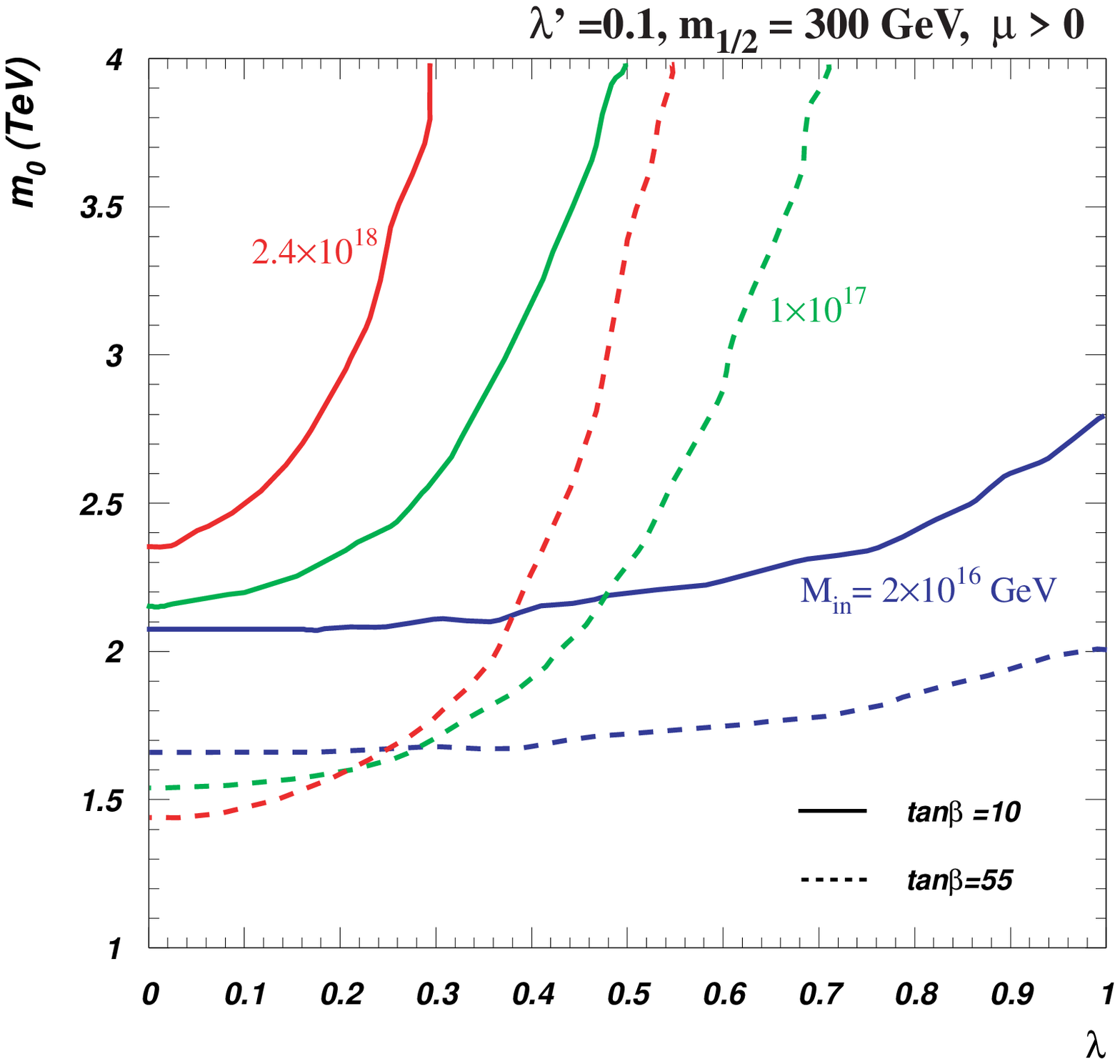,height=8.0cm}
\caption{\it
The value of $m_0$ at the boundary of electroweak symmetry breaking for $m_{1/2} = 300$~GeV,
$A_0 = 0$ and $\tan \beta = 10$ (solid) and =55 (dashed), as a function 
of $M_{in}$ for various values of $\lambda$ (left panel)
and as a function of $\lambda$ for various $M_{in}$ (right panel).}
\label{fig:MUm0}
\end{figure}

An example of how the value of $m_0$ at the
boundary of electroweak symmetry breaking depends on $M_{in}$
for different values of $\lambda$ is shown in the left panel of
Fig.~\ref{fig:MUm0}, where we have chosen
$m_{1/2} = 300$~GeV, $A_0 = 0$. We see from the solid curves that when $\tan \beta = 10$
the EWSB boundary value of $m_0$ increases monotonically with $M_{in}$ for any
value of $\lambda$, and that the rate of increase itself grows with the
value of $\lambda$. This is the result of the $\mu$ parameter getting larger (for fixed $m_0$)
with $M_{in}$,
as we discussed in the previous Section. 
However, for $\tan \beta=55$, shown by the dashed curves,
for small $\lambda$ the EWSB boundary migrates to smaller
$m_0$ as $M_{in}$ increases. For such large values of $\tan \beta$, 
$\hfiv \gg \hten$, producing a stronger downward
push in the evolution of
$m_{\calh_1}$ relative to that of $m_{\calh_2}$. If $\lambda \lesssim 0.3$, 
its effects in the RGE's  are negligible compared to those of the 
$h_{\mathbf{\overline{5}},\mathbf{{10}}}$ Yukawa
couplings, resulting in a hierarchy $m_{\calh_1}<m_{\calh_2}$ at $\mgut$.
In this case, we have $S>0$ at $\mgut$
which in turn leads
to smaller values of $\mu$ and hence a smaller value of $m_0$
at the electroweak symmetry breaking boundary. 
At the `critical' value of $\lambda \simeq 0.3$,
the evolutions of the $\calh_1$ and $\calh_2$
soft masses are almost identical, with the result that
at $\mgut$, $m_{\calh_1}\simeq m_{\calh_2} < m_0$,
so that $S \approx 0$ as in the CMSSM. 
As a result, the weak-scale value of $\mu$ is nearly the same as in the CMSSM -- the 
well-known focusing behaviour -- and the location
of the EWSB boundary remains stable with respect to $M_{in}$.
Finally, for larger $\lambda$, $S < 0$ at $\mgut$, 
the weak-scale value of $\mu$ is larger than in the CMSSM, and
the EWSB boundary migrates to larger $m_0$.

The effects of $\lambda$ are seen explicitly in the right
panel of Fig.~\ref{fig:MUm0}, 
where we show the location of the
EWSB boundary as a function of $\lambda$ for different choices of $M_{in}$ and two 
values of $\tan\beta$. We note
that even for vanishing $\lambda$ the EWSB boundary changes with $M_{in}$ due to the additional
running of the soft masses above $\mgut$.

\section{Summary of Results for Different \texorpdfstring{$\tan \beta$}{tanbeta}}
\label{sec:tanb}

As we have discussed above, the  $\stau$ coannihilation region tends
to disappear as $M_{in}$ increases, and the focus-point region tends to recede to larger $m_0$
as $\lambda$ increases. So far, we have shown these effects only for $\tan \beta = 10$ and 55.
Here we summarize how these effects vary for intermediate values of $\tan \beta$.

\begin{figure}[ht]
\epsfig{file=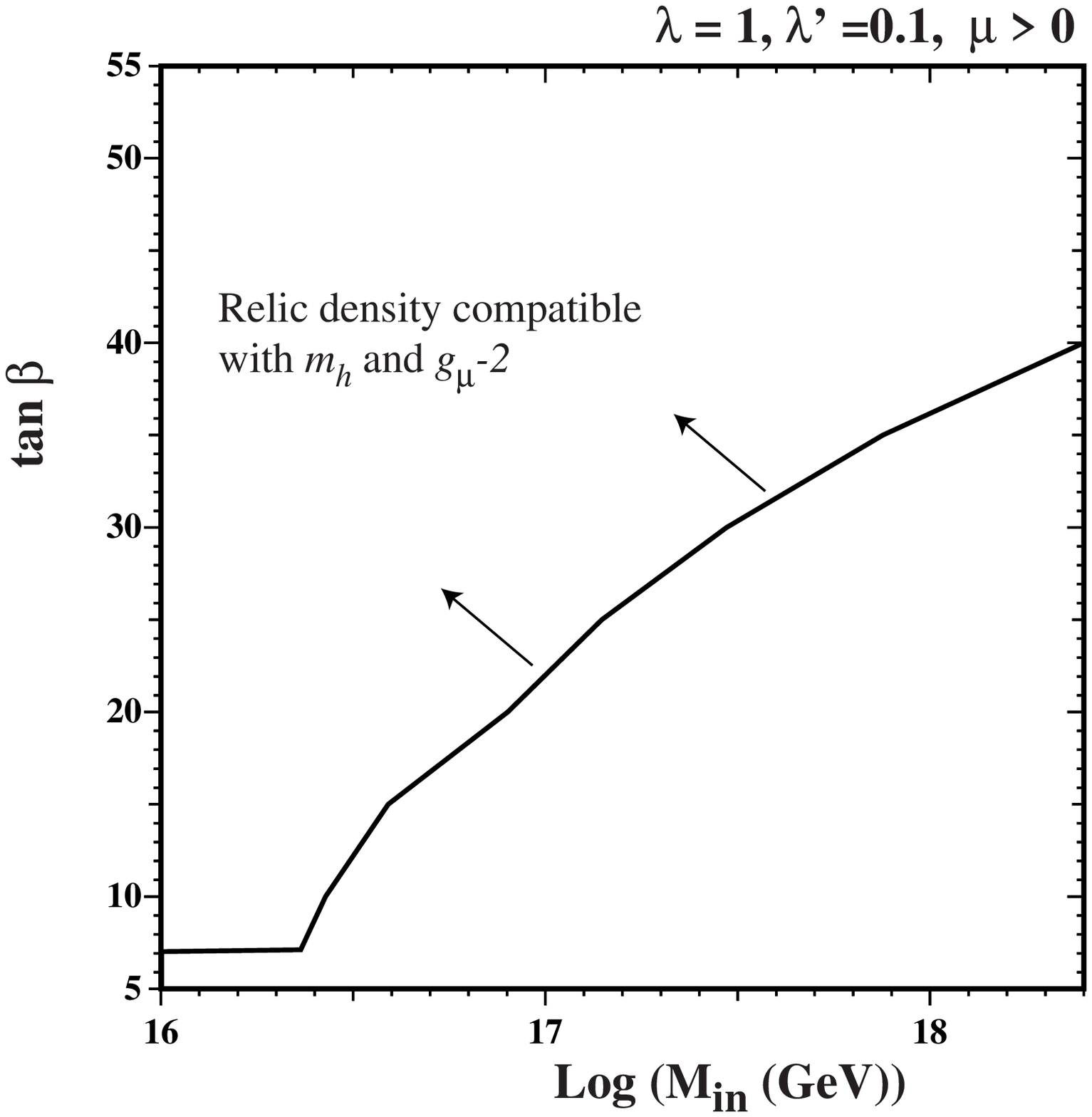,height=8.0cm}
\epsfig{file=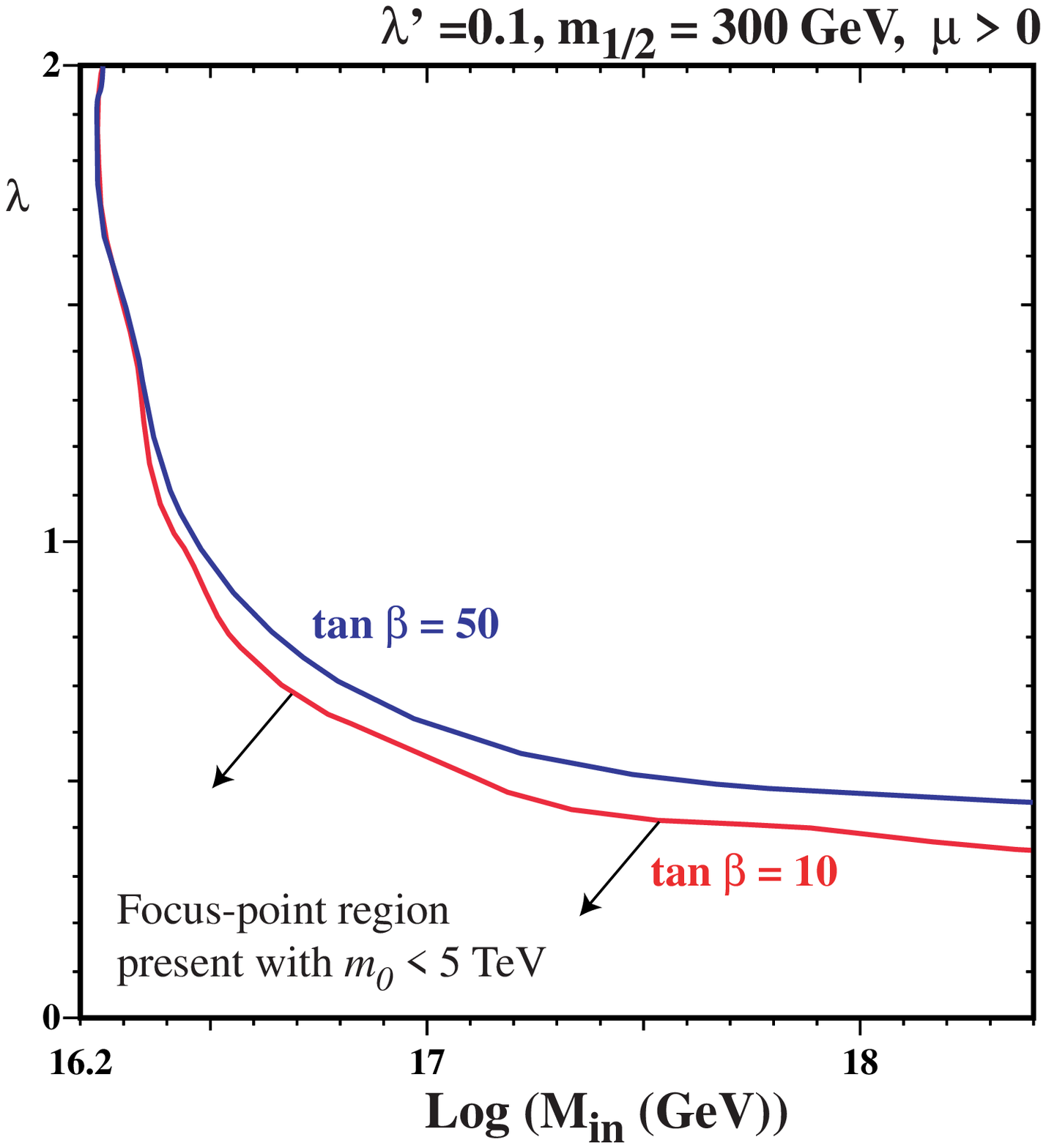,height=8.0cm}
\caption{\it
Left: the coannihilation/rapid-annihilation strips are compatible with other experimental
constraints only for values of $(M_{in}, \tan \beta)$ above the diagonal contour.
Right: the focus-point strip has $m_0 < 5$~TeV at $m_{1/2} = 300$~GeV only for
values of $M_{in}, \lambda$ below the red (blue) line for $\tan \beta = 10 (50)$.}
\label{fig:Summary}
\end{figure}

We see in the left panel of Fig.~\ref{fig:Summary} the region of the $(M_{in}, \tan \beta)$
plane where there is a coannihilation/rapid-annihilation strip. For choices lying below the
contour, all the points where coannihilation or rapid-annihilation brings the relic density
into the WMAP range are excluded by other constraints. For $\tan \beta \lappeq 20$,
the region below the 
curve has $m_h$ lower than the LEP bound, while for larger $\tan \beta$, $g_\mu -2$ is too large in this region~\footnote{The
kink in the boundary contour is an artifact of our approximation to the $m_h$ constraint:
incorporating the uncertainty in the theoretical calculation of $m_h$ and the
experimental likelihood function would smooth it out. Recall that for very low $\tan \beta$,
there is some tension between the Higgs mass and $g_\mu -2$ constraints.}. 
We see that, as
$\tan \beta$ increases above 10, the coannihilation/rapid-annihilation strip persists up to
progressively larger values of $M_{in}$. However, only for $\tan \beta > 47$ does it persist
for $M_{in} = \overline{M_P}$. This information is potentially a useful diagnostic tool, if
supersymmetry is discovered. For example, if experiments determine that $m_0$ is
(essentially) universal and $m_0 \ll m_{1/2}$ with $\tan \beta \sim 20$, then we can infer
from the left panel of Fig.~\ref{fig:Summary} that $M_{in} < 10^{17}$~GeV.

We have also seen earlier that the focus-point region is sensitive to the value of $\lambda$.
The right panel of Fig.~\ref{fig:Summary} displays the regions of the $(M_{in}, \lambda)$
plane where the focus-point strip has $m_0 < 5$~TeV for $m_{1/2} = 300$~GeV. We see
immediately that these regions are rather similar for $\tan \beta = 10$ and 50, lying below
the red and blue lines, respectively. This information could be used to infer a constraint
on $\lambda$, which otherwise does not impact significantly low-energy phenomenology.
For example, if experiment indicates that Nature is described by a focus-point model with 
$m_{1/2} = 300$~GeV, $m_0 < 5$~TeV and $M_{in} > 10^{17}$~GeV, then we can infer
from the right panel of Fig.~\ref{fig:Summary} that $\lambda < 0.6$.

\section{Summary}
\label{sec:concl}

We have shown in this paper that the characteristic $(m_{1/2}, m_0)$ planes
of the CMSSM are significantly modified as $M_{in}$ increases above $M_{GUT}$,
and also depend on the SU(5) GUT Higgs coupling $\lambda$, whereas
they are less sensitive to the other Higgs coupling $\lambda'$. Indeed, the familiar
stau coannihilation strip and focus-point region may disappear to small $m_{1/2}$
and large $m_0$, respectively, as $M_{in}$ increases. These possibilities should
be borne in mind when searching for supersymmetry at the LHC and elsewhere:
if Nature turns out to choose either the stau coannihilation strip or the focus-point
region, one may be able to derive an upper limit on $M_{in}$ and derive a
constraint on the GUT Higgs coupling $\lambda$.

These observations serve as another reminder that, although 
the CMSSM with its universal soft supersymmetry-breaking masses at the GUT
scale is appealingly simple, even small modifications of its assumptions may
change significantly the expected phenomenology. The CMSSM may be a comfortable
base camp for exploring the landscape of supersymmetric phenomenology,
but one must get out into the fresh air from time to time!

\section{Acknowledgements}
The work of A.M. and K.A.O. is supported in part by DOE grant DE-FG02-94ER-40823 at the 
University of Minnesota.

\end{document}